\documentclass[12pt]{article}
\pdfoutput=1
\usepackage[hyperfootnotes=false]{hyperref}
\usepackage[margin=.14\paperwidth]{geometry}
\usepackage{epsfig}
\usepackage{amsmath}
\usepackage{amssymb}
\usepackage{caption}

\numberwithin{equation}{section}

\usepackage{setspace}
\usepackage{graphicx}
\usepackage{color}
\usepackage{xcolor}

\usepackage{subcaption}
\usepackage{braket}

\def\be{\begin{equation}}
\def\ee{\end{equation}}
\def\ba#1\ea{\begin{align}#1\end{align}}
\def\bg#1\eg{\begin{gather}#1\end{gather}}
\def\bm#1\em{\begin{multline}#1\end{multline}}
\def\bmd#1\emd{\begin{multlined}#1\end{multlined}}

\def\({\left(}
\def\){\right)}
\def\[{\left[}
\def\]{\right]}
\def\<{\langle}
\def\>{\rangle}

\def\tr{\operatorname{tr}}

\def\nref#1{(\ref{#1})}

\begin{document}

\begin{titlepage}
  \bigskip

  \bigskip\bigskip\bigskip\bigskip

  \bigskip

\centerline{\Large \bf { Modular Flow as a Disentangler }}

  \bigskip

  \begin{center}

\bf {Yiming Chen$^1$, Xi Dong$^2$, Aitor Lewkowycz$^3$, Xiao-Liang Qi$^{3,4}$}

  \bigskip \rm
  \bigskip

$^1${\it Department of Physics, Tsinghua University, Beijing, 100084, China}

  \smallskip
 
$^2${\it Department of Physics, University of California, Santa Barbara, CA 93106, USA}

  \smallskip

$^3${\it Stanford Institute for Theoretical Physics, Department of Physics,\\ Stanford University, Stanford, CA 94305, USA}

  \smallskip

$^4${\it School of Natural Sciences, Institute for Advanced Study, Princeton NJ 08540, USA}

  \smallskip 
 
  \vspace{1cm}
  \end{center}

  \bigskip\bigskip

  \bigskip\bigskip

\begin{abstract}

In holographic duality, the entanglement entropy of a boundary region is proposed to be dual to the area of an extremal codimension-2 surface that is homologous to the boundary region, known as the Hubeny-Rangamani-Takayanagi (HRT) surface. 
In this paper, we study when the HRT surfaces of two boundary subregions $R,A$ are in the same Cauchy slice. This condition is necessary for the subregion-subregion mapping to be local for both subregions and for states to have a tensor network description. To quantify this, we study the area of a surface that is homologous to $A$ and is extremal except at possible intersections with the HRT surface of $R$ (minimizing over all such possible surfaces), which we call the constrained area. We give a boundary proposal for an upper bound of this quantity, a bound which is saturated when the constrained surface intersects the HRT surface of $R$ at a constant angle. This boundary quantity is the minimum entropy of region $A$ in a modular evolved state -- a state that has been evolved unitarily with the modular Hamiltonian of $R$. We can prove this formula in two boundary dimensions or when the modular Hamiltonian is local. This modular minimal entropy is a boundary quantity that probes bulk causality and, from this quantity, we can extract whether two HRT surfaces are in the future or past of each other. These entropies satisfy some inequalities reminiscent of strong subadditivity and can be used to remove certain corner divergences.  

\end{abstract}

\end{titlepage}

\tableofcontents

\section{Motivation}

In holographic duality \cite{Maldacena:1997re,Gubser:1998bc,Witten:1998qj}, the Hubeny-Rangamani-Ryu-Takayanagi (HRT) formula  \cite{Ryu:2006bv,Ryu:2006ef,Hubeny:2007xt} 
\begin{equation}
S_A=\frac{|\gamma(A)|}{4G_N}
\end{equation}
relates the entanglement entropy of a boundary region $A$ to the area of the extremal surface $\gamma(A)$ that is homologous to $A$. The HRT formula was originally proposed for geometries with time translation symmetry, where extremality implies minimality on the preferred Cauchy slice in the bulk.  The extremal surface, known as the HRT surface, can also be defined by a maximin procedure \cite{Wall:2012uf}: one can first find the minimal surface $\gamma(A)|_{\Sigma} \in \Sigma$ in a given Cauchy surface $\Sigma$ which includes region $A$ in its intersection at the boundary. Then the actual HRT surface is obtained by varying $\Sigma$ and find the maximum of the area of $\gamma(A)|_{\Sigma}$. In this procedure it is clear that the extremal surface is always a saddle surface, the area of which increases upon variations along space-like directions, and decreases upon variations along time-like direction. 

\begin{figure}[h]
\centering
\includegraphics[width=4in]{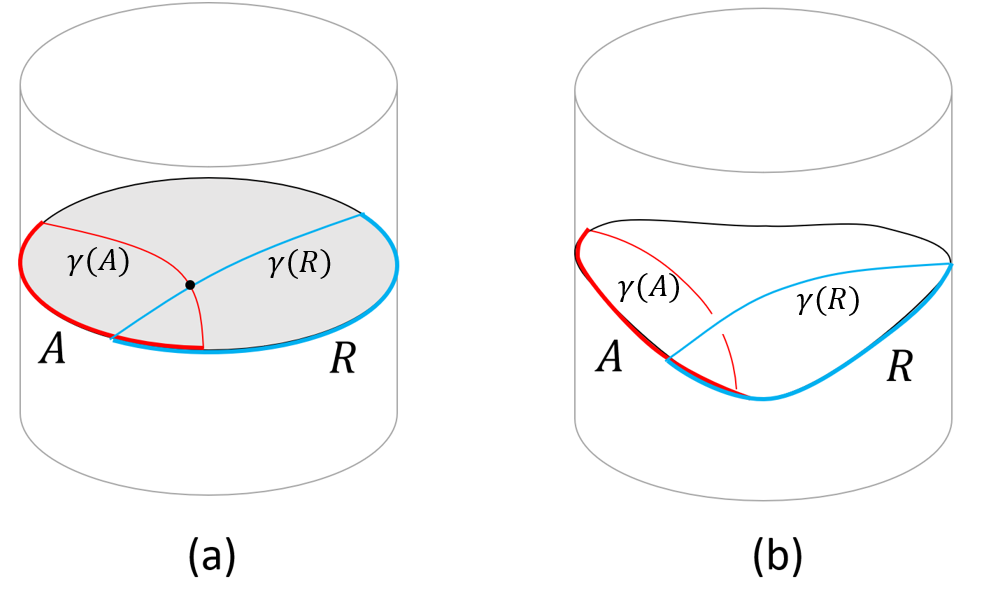}
\caption{Illustration of HRT surfaces for different boundary regions. (a) For a state with time reflection symmetry, all HRT surfaces of different boundary regions lie in the same bulk slice, so that HRT surfaces of boundary overlapping regions must intersect. (b) For a generic boundary state defined by a boundary Cauchy slice, the HRT surfaces of two overlapping regions $R$ and $A$ generically cannot be included in any bulk Cauchy slice, if they are time-like separated. }
\label{fig:setup}
\end{figure}

The HRT formula uncovers an  intrinsic connection between spacetime geometry and quantum entanglement. In gravity, there is no fundamental meaning to any particular bulk slice: different slices are all gauge equivalent. However, the HRT surface corresponding to a boundary subregion lives in a proper subset of all possible gauge equivalent bulk slices. This is a surprising property which suggests that when focusing on boundary subregions, not all bulk Cauchy slices are equally preferred. As has been discussed in \cite{Headrick:2014cta}, the HRT surface defines the entanglement wedge $a$ as the bulk domain of dependence of an achronal bulk surface (known as a homology hypersurface) whose boundary is $A\cup\gamma(A)$. The entanglement wedge plays an important role in subregion-subregion duality:  the algebra of bulk operators localized in the entanglement wedge $a$ is encoded
 in the boundary region $A$ \cite{Almheiri:2014lwa,Dong:2016eik}. When all HRT surfaces lie in the same bulk Cauchy slice, there is a preferred bulk slice where this mapping between bulk and boundary subregions acts locally (Fig. \ref{fig:setup} (a)). However, generically, we expect that for two {overlapping} boundary subregions $R,A$, their HRT surfaces are not in the same bulk Cauchy slice ({\it i.e.} they are time-like separated, see Fig. \ref{fig:setup}(b)). In this situation, there cannot be a local description for both bulk algebras of operators in the same Cauchy slice. In other words, the mapping from boundary operators in a region to bulk local operators has to be different for $R$ and $A$. This is reminiscent of the code subspace story of \cite{Almheiri:2014lwa}.  The reason why reconstruction of bulk operators on the boundary is possible is ultimately that bulk and boundary modular flows are the same \cite{Jafferis:2015del,Dong:2016eik,Faulkner:2017vdd} and, in this paper, we will also see how this property of modular flow can be used to define a boundary quantity that quantifies whether two HRT surfaces can be put in the same bulk slice.

An important motivation of studying whether HRT surfaces of different regions lie in one bulk Cauchy slice comes from the tensor network picture. Since B. Swingle's work \cite{Swingle:2009bg} there have been various proposals relating tensor networks to holographic duality \cite{nozaki2012holographic,qi2013exact,Pastawski:2015qua,Yang:2015uoa,Hayden:2016cfa}. Tensor networks are representations of many-body quantum states by contracting tensors on a given graph. Some classes of tensor networks, such as random tensor networks with large bond dimension in Ref. \cite{Hayden:2016cfa} satisfy the RT formula for the entanglement entropy of any boundary region, where the area of a surface is defined as number of links intersecting the surface. Therefore if we consider a state in a holographic theory for which all HRT surfaces lie in a single bulk Cauchy slice, it is natural to compare it with a tensor network state with a graph geometry that is obtained by discretizing the particular Cauchy slice. For more generic states, the HRT surfaces for two intersecting regions on the boundary do not intersect in the bulk, and therefore there is no natural choice of Cauchy slice for considering a tensor network representation.\footnote{More precisely, one can always describe such a state with a tensor network, but the geometrical entropy upper bound will not be saturated for at least one of the two regions considered.} The difficulty in a tensor network representation of such states suggests that 
there shall be a {\it quantum information measure} of the boundary state which probes whether the HRT surfaces for two regions in a holographic state intersect or not. Finding such quantum information measure will help improve our understanding on the relation of bulk dynamics and boundary entanglement structure.

To look for a measure of this property, we first quantify the difference between intersecting and non-intersecting HRT surfaces by defining a constrained extremal surface ${\gamma}_R(A)$, which is homologous to a region $A$ and is allowed to intersect the HRT surface $\gamma(R)$ of the other region $R$, if this can reduce its area. The area difference between the constrained surface and the actual HRT surface, $|\gamma(A)|-|{\gamma}_R(A)|$, is a measure of how far away (in time) $\gamma(A)$ and $\gamma(R)$ are from each other. We propose a boundary dual of this difference, which is an entropy reduction by modular flow. More precisely, when we have constrained surfaces that intersect $\gamma(R)$ at a constant boost angle we will have a precise boundary quantity that equals the area of $\gamma_R(A)$, and more generally, it will be an upper bound. We will provide more details of the definition later, but the basic idea is that by modular evolving the state with respect to region $R$, i.e. by applying a unitary operator that is defined as $\rho_R^{is}$ to the state, the entropy of region $A$ can be reduced, and the minimal entropy obtained by varying the modular flow time $s$ is proposed to be dual to the constrained extremal surface area (divided by $4G_N$). 

The remainder of the paper is organized as follows: in Section $2$, we elaborate on the constrained area and its conjectured boundary dual---the modular minimal entropy. Section $3$ exposes the evidence for this proposal. In Section $4$, we give some examples and applications of the formalism. We conclude with Section $5$, where we comment on possible extensions and further applications of our results. 

During the completion of this work, we became aware of \cite{Faulkner:2018faa} which has some partial overlap with the results of this paper.

\section{Proposal}

To begin with, we would like to propose a bulk quantity which quantifies whether two HRT surfaces, $\gamma(R), \gamma(A)$ (corresponding to boundary regions $R,A$ on a boundary Cauchy slice) can be in the same bulk slice. As shown in \cite{Wall:2012uf}, if $R \in A$ (or $R \in \bar{A}$), this is always possible (entanglement wedge nesting). The non-trivial case is then when $R, A$ have some partial overlap. In this case, if the HRT surfaces do not intersect, they necessarily cannot be put in the same bulk slice. In the rest of the paper we will say that two codimension-$2$ surfaces are space-like separated when they can be put in the same Cauchy slice and time-like separated when there does not exist a Cauchy slice which contains both surfaces. In this way, we want to consider a new geometric object: the constrained extremal surface, defined as the bulk extremal surface which is homologous to $A$ and can intersect  $\gamma(R)$ (as long as this minimizes the area):
\begin{equation}
   \gamma_R(A) \equiv \text{min}_{|\gamma|}  \lbrace \gamma \text{ is extremal except across } \gamma \cap \gamma(R) , \partial \gamma= \partial A \rbrace\label{eq:cond}
\end{equation}

\begin{figure}[h]
\centering
\includegraphics[width=4in]{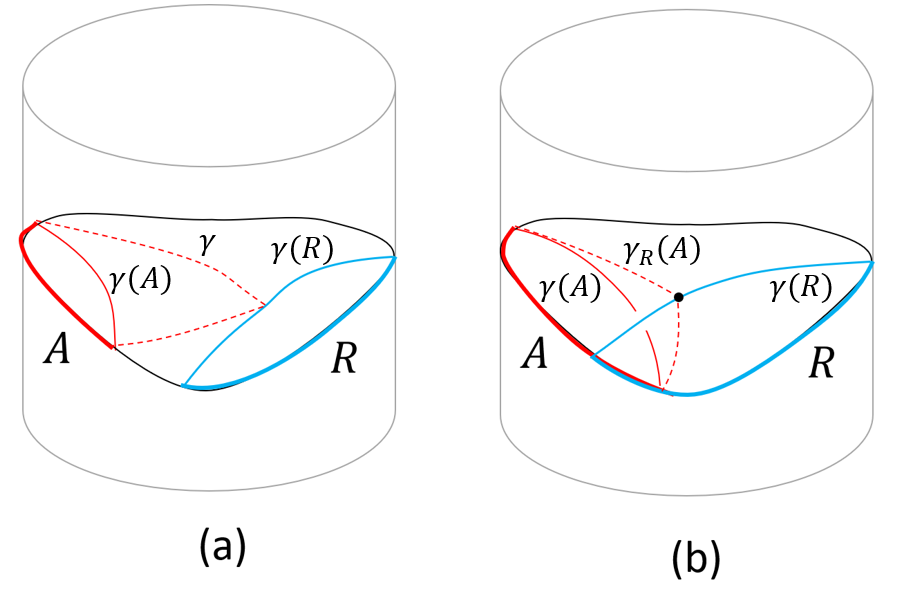}
\caption{Illustration of the definition of constrained surface $\gamma_R(A)$. (a) When $A$ and $R$ do not overlap, it is always possible to fit $\gamma(A)$ and $\gamma(R)$ in the same Cauchy slice, so that any constrained surface $\gamma$ (red dashed curve) that is extremal everywhere except for the intersection with $\gamma(R)$ will have an area bigger than $\gamma(A)$. (b) When $A$ and $R$ overlap, it is possible to have a constrained surface with smaller area than $\gamma(A)$, in which case this surface is $\gamma_R(A)$ (see text). }
\label{fig:def}
\end{figure}

In other words, the constrained extremal surface $\gamma_R(A)$ has zero trace of the extrinsic curvature everywhere in the bulk expect at possible intersections with $\gamma(R)$.  It is not required to intersect with $\gamma(R)$, and in particular the original extremal surface $\gamma(A)$ also satisfies the condition in Eq.~(\ref{eq:cond}).  Among all surfaces satisfying the condition in Eq.~(\ref{eq:cond}), the constrained extremal surface is the one with the smallest area.
This definition is illustrated in Fig.~\ref{fig:def}. If $\gamma(A),\gamma(R)$ are space-like separated, they can be put in the same Cauchy slice, so that any constrained surface $\gamma$ that satisfies Eq. (\ref{eq:cond}) and intersects $\gamma(R)$ must have a bigger area than $\gamma(A)$. Therefore in that case $\gamma_R(A)=\gamma(A)$. In contrast, if $\gamma(A)$ and $\gamma(R)$ are time-like separated, there exists a constrained surface with nontrivial intersection with $\gamma(R)$ with a smaller area than $\gamma(A)$.  The intersection is generically a codimension-3 surface in the bulk.

We could also give a maximin definition as in \cite{Wall:2012uf}, where the Cauchy slices are forced to contain $\gamma(R)$, but we find the above definition better because it is more local. 
By construction,  $|\gamma(A)|-|\gamma_R(A)| \ge 0$, and the inequality saturates if $\gamma(A) , \gamma(R)$ are spacelike separated. 

\begin{figure}[h]
\centering
\includegraphics[width=4in]{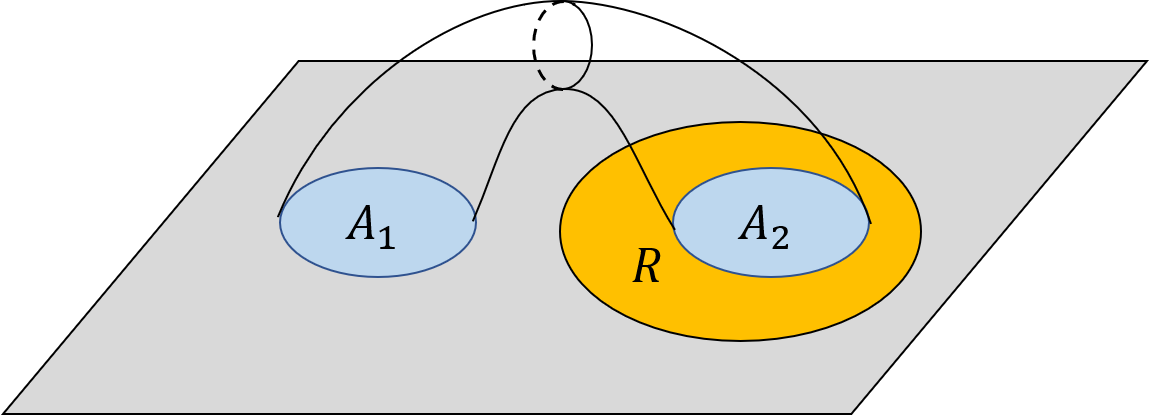}
\caption{Example of a situation with $\partial A \cap \partial R=\emptyset$ when the boundary dimension is $(2+1)$. The HRT surface of $A=A_1\cup A_2$ is also drawn schematically.}
\label{fig:higherdim}
\end{figure}

We expect that the constrained surface is non-trivial even if $\partial A \cap \partial R$ is non-empty (which is generic in more than $2$ dimensions). Even if these surfaces intersect in the boundary, they generically will not have additional intersections in the bulk. As we will discuss later, the difference between the constrained area and the original area in this case can be divergent.  In higher dimensions, we  can also have $\partial A \cap \partial R=\emptyset$ and, in these situations, {we do not expect the divergence structure to change}: some examples of these higher dimensional situations are two strips or $A$ being two spheres $A_1,A_2$ and $R$ a bigger sphere surrounding $A_1$ (See. Fig. \ref{fig:higherdim}). 

In order to have a precise boundary dual, we would further want to restrict to the constrained surfaces that intersect $\gamma(R)$ at a constant boost angle:
\begin{equation}
    \gamma_R^{>}(A)= \text{min}_{|\gamma|}  \lbrace \gamma \text{  is extremal except having constant boost angle across } \gamma \cap \gamma(R) , \partial \gamma= \partial A \rbrace
\end{equation}

 For a generic $\gamma(R)$, at any point where $\gamma_R(A)$ intersects the surface, there will be an incoming and an outgoing vector. The boost angle is defined by the inner product of the projection of these vectors to the normal plane of $\gamma(R)$ at that point. In $d=2$ for one interval, the constrained surfaces will all be at a constant boost angle. For multiple regions in $d=2$ and higher dimensions, it is not always guaranteed that there exists such a surface. 

In the boundary, we expect that there is some quantum information quantity that computes the area of the constrained surface, which we will call modular minimal entropy: $\bar{S}_R(A)$.  Given our state $|\Psi \rangle$ and $\rho_R$, we can use modular evolution (evolution with the logarithm of the density matrix) to obtain a one parameter family of states:
\begin{equation}
    |\Psi(s_R) \rangle \equiv \rho_R^{i s_R} |\Psi \rangle
\end{equation}
{Here $s_R$ is real, so that $\rho_R^{is_R}$ is unitary.}

For this family of states, we can compute the entanglement entropy of $A$, $S_A(\rho(s_R))$ and, in general, this entropy can be bigger or smaller than the original entropy (at $s_R=0$). Our proposal is that the minimum of this object with respect to $s_R$, which we call the modular minimal entropy, is precisely the area of $\gamma_R^> (A)$ divided by $4G_N$:

\begin{equation}
    \bar{S}_R(A) \equiv \text{min}_{s_R} S_A(\rho(s_R))=\frac{|\gamma^{>}_R(A)|}{4 G_N} \label{Smin}
\end{equation}

This boundary definition of the modular minimal entropy trivially satisfies the condition $S(A)-\bar{S}_R(A) \ge 0$. In the next subsection we will discuss how it is derived for two-dimensional boundary theories and for the cases when the modular Hamiltonian is local.

In the situations where there does not exist any non-trivial $\gamma^{>}_R(A)$, we do not expect $\bar{S}_R(A)$ to have a bulk interpretation in the original geometry. However, as we will show in $d=2$ (for multiple regions) and we conjecture for higher dimensions, we generally expect:
\begin{equation}
   S(A)\ge \bar{S}_R(A) \ge \frac{|\gamma_R(A)|}{4 G_N}
\end{equation}
where $\gamma_R(A)$ is the minimal constrained surface which does not necessarily intersect $\gamma(R)$ with a constant boost angle. Because of this,   $S(A)-\bar{S}_R(A)$ is still a good diagnostic of whether the surfaces intersect: when they intersect, all inequalities will be saturated, and we will necessarily have $S(A)=\bar{S}_R(A)=\frac{|\gamma_R(A)|}{4 G_N}$.

A consistency check of this formula is the case where there is a $Z_2$ time reflection symmetry. In this case, in the bulk, the two surfaces will be in the same slice and thus the minimum will be at zero modular parameter $s_{min,R}=0$. In the boundary, we can check that $s_R=0$ is an extremum: around $s=0$, we can use the first law of entanglement entropy to derive $\partial_s S_A(\rho_R(s))|_{s=0}=i \langle [K_R,K_A] \rangle$. {The operator on the right-hand side is odd in time reflection symmetry, and thus has to vanish in a reflection symmetric state.}

\section{Evidence}

In this section, we will {discuss two cases when proposal \nref{Smin} can be verified. The first case is in general dimension, when the modular Hamiltonian $-\log \rho_R$ is local. The second case is for two-dimensional boundary theory with arbitrary regions (and states).}

\subsection{Local modular Hamiltonians}

When the modular Hamiltonian is a local integral of the (CFT) stress tensor, modular evolution is just Hamiltonian evolution and we can understand $|\psi (s_R)\rangle$ explicitly. Two known situations where this happens are when {$R$ is} a spherical subregion in the vacuum of a CFT (or the half-plane) or one CFT in the thermofield double state (TFD). In these two cases, despite their simplicity we can get a non-trivial $\bar{S}_R(A) \not = S(A)$. 

Consider the time-evolved TFD state:
\begin{equation}
    |TFD(t) \rangle = \sum_i e^{-(\frac{\beta}{2}+i t) E_i} |E_{i,L},E_{i,R} \rangle
\end{equation}
The entropy of the right CFT is time independent and given by the thermal entropy $S_R(t)=S(\beta)$. We would like to consider modular flow with respect to region $R$, which in this case just corresponds to right time evolution: $e^{-i K_R s_R}|TFD(t)\rangle=|TFD(t+s_R ) \rangle$. 

\begin{figure*}[h!]
    \centering
    \begin{subfigure}[b]{0.3\textwidth}
     \centering
     \includegraphics[height=1.3in]{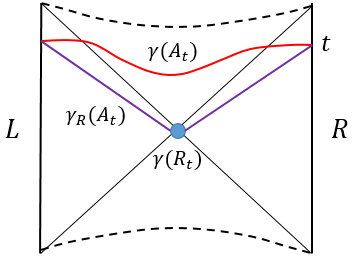}
        \caption{}
    \end{subfigure}%
    ~ 
        \begin{subfigure}[b]{0.3\textwidth}
     \centering
     \includegraphics[height=1.3in]{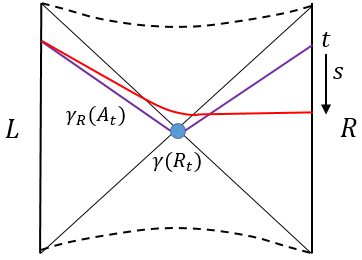}
        \caption{}
    \end{subfigure}%
    ~ 
    \begin{subfigure}[b]{0.3\textwidth}
        \centering
        \includegraphics[height=1.3in]{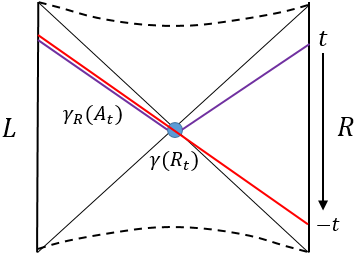}
        \caption{}
    \end{subfigure}
    \caption{(a) We consider the situation of the $\ket{TFD(t)}$ state, where $\gamma(A_t),\gamma(R_t)$ are not in the same Cauchy slice. The constrained surface $\gamma_R (A_t)$ crosses the bifurcation surface. (b) Modular evolution shifts the endpoint of the HRT surface $\gamma (A_t)$. The entanglement entropy $S(A)$ after the modular evolution can be calculated through the length of the shifted HRT surface. (c) The shifted HRT surface reaches minimal length when it goes through the bifurcation horizon $\gamma(R_t)$. By symmetry, the area of this surface is the same as that of $\gamma_R(A_t)$.}
    \label{fig:localTFD}
\end{figure*}

We would like to define the modular minimal entropy for the union of two half-planes on the left and right CFT: $A=A_L \cup A_R$. First, in the $\ket{TFD(t)}$ state, the corresponding HRT surface $\gamma(A_t)$ is time dependent and goes through the interior of the black hole \cite{Hartman:2013qma}. This surface is clearly not in the same Cauchy slice as the $\gamma(R_t)$ which is the bifurcation horizon (see figure \ref{fig:localTFD}). In this case, the modular minimal entropy is given by the area of the surface that ends in $\partial A_t$ and goes through $\gamma(R_t)$. Because of the symmetries of the problem, this surface has the same area as the HRT surface that goes between $\partial A_{L,t}$ and $\partial A_{R,-t}$. Due to boost invariance, this area is independent of $t$, so that $|\gamma_R(A_t)|=|\gamma(A_{t=0})|$. On the other hand, the minimization over modular flow clearly happens when $s_R=-2t$, so $\bar{S}_R(A_t)=S(A_{t=0})$ which coincides with the area of the constrained surface (see figure \ref{fig:localTFD} for more details). In other words, in this case the modular flow minimizes entropy by ``undoing" the time evolution of the TFD state.

\begin{figure*}[h!]
    \centering
        \begin{subfigure}[b]{0.5\textwidth}
     \centering
     \includegraphics[height=2.5in]{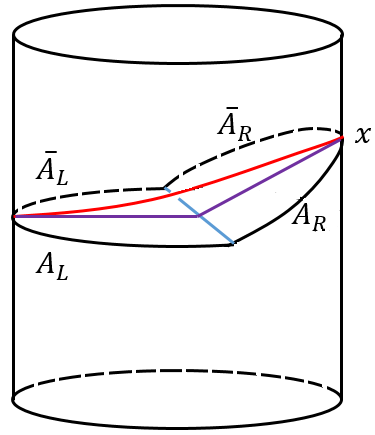}
        \caption{}
    \end{subfigure}%
    ~ 
    \begin{subfigure}[b]{0.5\textwidth}
        \centering
        \includegraphics[height=2.5in]{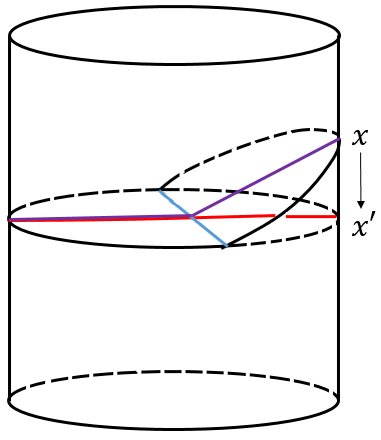}
        \caption{}
    \end{subfigure}
    \caption{The sphere in the vacuum situation is similar. In the figures, we show how it works in AdS$_3$. (a) The surfaces $\gamma (A)$, $\gamma (R)$ and $\gamma_R (A)$ are in color red, blue and purple, respectively. (b) The role of the modular flow is to move the right endpoint of $A$, $x$. The minimum is obtained when $x$ is shifted to $x'$ in the $t=0$ slice. Because of the symmetries, the purple and red surfaces have the same area.
    }
    \label{fig:localvac}
\end{figure*}

For spheres in the vacuum, the situation is pretty much the same. We want to divide our system into four regions: $A_L,\bar{A}_L,A_R,\bar{A}_R$. The vacuum state in the original $t=0$ surface does not have any interesting dynamics, but we can consider more time dependent slice: $e^{-i K_R t}|0 \rangle$ (or a more regular version of this). The two regions of interest are defined as $A=A_L\cup A_R$, $R=A_R\cup \bar{A}_R$. The details do not matter too much as long as $\partial R$ is held fixed and $\partial A$ is not in the time reflection symmetric $t=0$ slice. In this case, $\gamma(R),\gamma(A)$ will not be in the same slice and thus $\gamma_R(A)$ will be non-trivial. As in the TFD case, we can think of the modular flow $s_R$ as moving the right endpoint of $A$, and the minimum will be obtained when it aligns with the left endpoint of $A$ (their HRT surface goes through $\gamma(R)$). Then, because of symmetry this entropy will be the same as $\gamma_R(A)$ and will be the same as the entropy of $A$ in the $t=0$ surface (see figure \ref{fig:localvac}).

\subsection{Non-local modular flows for 2-dimensional holographic theories}

Beyond the local case, modular evolution will be non-local and rather complicated. Generally, we do not know how to think about the bulk dual of the modular flowed state, 
but as we will explain below, in two boundary dimensions we can still make some progress. 

Let us start from the formula for the integral over modular flow for local heavy operators of dimension $c \gg \Delta \gg 1$ from \cite{Faulkner:2017vdd}:
\begin{equation}
\int_{-\infty}^{\infty} ds_R    \langle  O(x_L) \rho_R^{-i s_R} O(y_R) \rho_R^{i s_R} \rangle_{\psi}= \text{max}_{z \in \gamma(R)} e^{-\Delta [d(x,z)+d(z,y)]} \label{zeromodes}
\end{equation}
This formula relates the integral over modular flow of the correlator of two local operators in $L,R$ with a geometric quantity: $d(x,z^*)+d(z^*,y)$ is the minimal geodesic distance between the two points $x,y$ with the constraint that it has to go through $\gamma(R)$, the HRT surface dual to region $R$. In other words, the right hand side consists of the sum of two geodesic distances between the boundary and the same bulk point in the HRT surface, which is chosen in such a way that the total distance is minimized.  The reason why this formula is possible is that bulk and boundary modular flow are equivalent \cite{Jafferis:2015del} and modular evolution close to the HRT surface is constrained by locality. Since we expect the modular evolved two point function to be exponentially suppressed in the mass of the heavy particle, we also expect that, in this limit, the LHS is dominated by some $s_{R,min}$, which maximizes the value of the correlator:

\begin{equation}
\int_{-\infty}^{\infty} ds_R    \langle  O(x_L) \rho_R^{-i s_R} O(y_R) \rho_R^{i s_R} \rangle_{\psi} \approx \langle  O(x_L)  O(y_R,s_{R,min})  \rangle_{\psi}  \label{zeromodesg}
\end{equation}

We are interested in the previous quantity because the R\'{e}nyi entropies $\tr \rho^n$ are given by the two point function of twist operators with dimension $\Delta_n=\frac{c}{12 n^2} (n^2-1) $. The entanglement entropy is obtained by taking one derivative of the $n \rightarrow 1$ limit of the twist operators. A large $c$ and $n \sim 1$, the dimension of the operator will be $O(c)$ but satisfies $\Delta_n/c \sim 0$ , so we can think of the twist operator as heavy but not backreacting operator \cite{Hartman:2013mia}. In this way, we can obtain $\bar{S}_{R}(A)$ from:
\ba
\bar{S}_{R}(A) &= -\partial_n \int_{-\infty}^{\infty} ds_R e^{-c (n-1) S_n(A,s_R)}|_{n=1}\\
&= -\partial_n \int_{-\infty}^{\infty} ds_R \langle T_n({\partial A_L}) e^{i K_R s_R} T_n(\partial A_R) e^{-i K_R s_R} \rangle|_{n=1}
\ea
where the first equality follows from taking the saddle point in the $s$-integral, which makes it localize at a particular $s$ (from $c \gg c(n-1) \gg 1$). Then, since to leading order in $(n-1)$ we can treat the twist operators as heavy but not backreacting local operators, we can apply \nref{zeromodes} to obtain\footnote{Using this expression is certainly justified when acting on states in the code subspace of low energy states.  Since modular flow acts in the code subspace, this approximation is well justified. } :
\begin{equation}
    \bar{S}_R(A)=\frac{c}{6}|\gamma_R(A)|
\end{equation}

It should be noted that whenever $\partial A \cap R=\emptyset$, by definition the entropy does not depend on the modular flow (since the effect of the modular flow is only nontrivial when one of the two operators experiences modular evolution). In the bulk, this is the statement that when one of these constrained surfaces enters and leaves a disconnected component of an entanglement wedge, there is no constraint in that region. 
Correspondingly, in the bulk if the entanglement wedge of $R$ has multiple disconnected components, the constraint only depends on the component that has nontrivial overlap with $\partial A$, {\it i.e.} the ones where the constrained surface $\gamma_R(A)$ ends at the boundary.

\subsection{An example with no constant-boost-angle constrained surface}

\begin{figure}[h]
\centering
\includegraphics[width=3in]{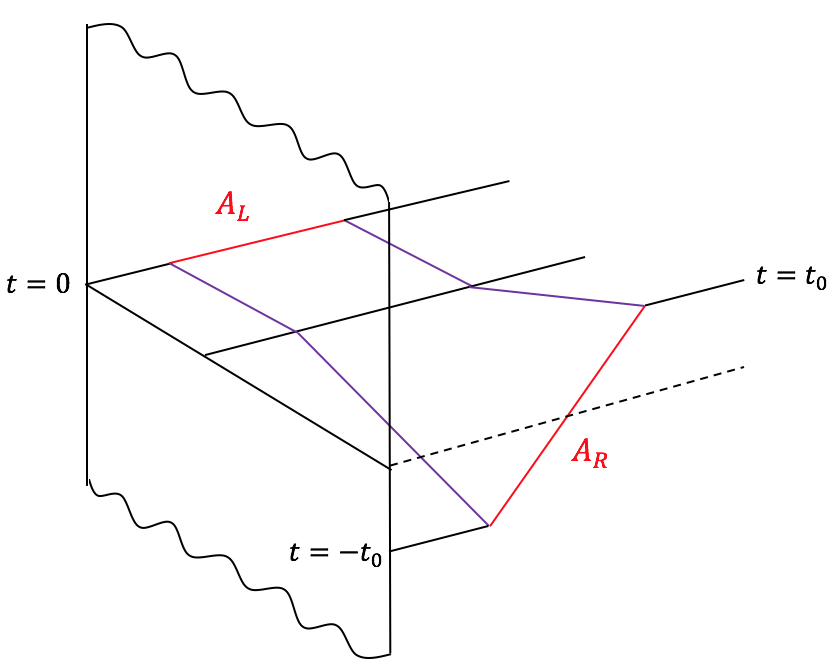}
\caption{Situation when the constraint surface has two different boost angles and thus it is different from $\bar{S}_R(A)$.}
\label{fig:tfdsign}
\end{figure}

In general, it may not be possible to find a constrained surface with constant boost angle at the intersection. This occurs generically when $\gamma(A)$ consists of multiple disconnected surfaces. As an example, consider again the $d=2$ case of the TFD state, but now with $R$ being the right CFT, $A_L$ being an interval at $t=0$ in the left CFT and $A_R$ a boosted interval, whose endpoints are at $t=\pm t_0$ respectively (See Fig. \ref{fig:tfdsign}). $\gamma(A)$ will consist of two disconnected surfaces in this case: one will be in the future of $\gamma(R)$ and the other in its past. The constraint surface $\gamma_R(A)$ will consist of two disconnected geodesics with different boost angles. Because of the setup, there is no constant boost angle that can reach $A_R$ from $\gamma(R)$ and thus $\gamma^{>}_R(A)$ does not exist. 

In this situation, since the constrained surface $\gamma_R(A)$ has two different boost angles, it can be thought as minimizing the object
\begin{equation}
    \langle T_n(\partial A_{L,1}) e^{i K_R s_1} T_n(\partial A_{R,1}) e^{-i K_R s_1} \rangle \langle T_n(\partial A_{L,2}) e^{i K_R s_2} T_n(\partial A_{R,2}) e^{-i K_R s_2} \rangle
\end{equation}
independently with respect to $s_1,s_2$. When $s_1  = s_2$, this is $S_A(\rho(s_1))$, but when they are different, it does not have a quantum information interpretation. Since this minimization is less constrained than $\bar{S}_R(A)$ it will necessarily be smaller:

\begin{equation}
    \bar{S}_R(A)> \frac{|\gamma_R(A)|}{4 G_N}
\end{equation}

More generally, for multiply disconnected regions, we expect each disconnected component of the modular minimal entropy will be determined by a different local boost parameter, which as discussed in \cite{Faulkner:2018faa} corresponds to a different value for $s$, even in the non-local setup. So, we expect that $\bar{S}_R(A)$ is generically an upper bound for the constrained area.

\subsection{Divergence structure in higher dimensions}

\begin{figure}[h]
\centering
\includegraphics[width=5in]{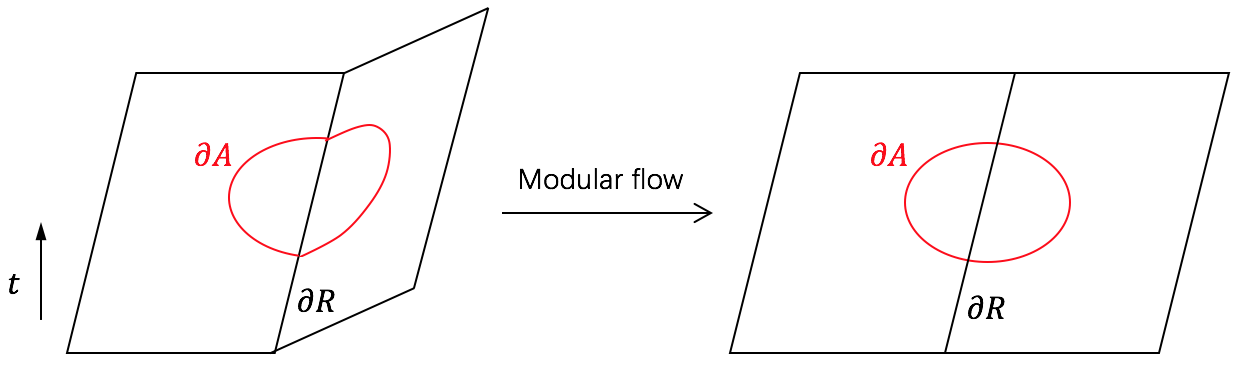}
\caption{When $\partial A\cap \partial R \not = \emptyset$, the modular flow can change the divergent structure. In this example, we see how the modular flow can get rid of the kink of a half-boosted sphere.  }
\label{fig:boosted}
\end{figure}

When the boundary dimension is two, the boundary of the two regions $\partial A$ and $\partial R$ are isolated points. In higher dimensions, the boundary of the two regions can have nontrivial intersection. For example Fig. \ref{fig:boosted} shows an example with $(2+1)$-d boundary. 
In that case, modular evolution of region $R$ introduces a kink at $\partial R$. From our definition of the modular minimal entropy, if  $\partial A \cap \partial R \not = \emptyset$, the modular evolved state will have this local kink.  This means that this constrained surface can change the structure of UV divergence. For example, if we have a half boosted sphere in flat space and we evolve with modular flow, the modular minimal entropy will maximally reduce the entropy of the sphere (see figure \ref{fig:boosted}), getting rid of the corner term divergence. Whether modular flow introduces or removes this kink depends on the sign of the divergence. Space-like corner contributions to entanglement entropy have been the subject of extensive study  \cite{Myers:2012vs,Bueno:2015rda,Bueno:2015xda}, but to our knowledge, the case where the corner angle is a boost has not been studied. Note that because this kink happens near $\partial R$ (which is kept fixed under modular evolution), the change in the divergence structure of the entropy of $A$ is universal: independent on the state or whether the modular flow is non-local. 
While in the simplest situation $\gamma_R(A)$ only intersects $\gamma(R)$ once, there can be more general situations with several connected components of the entanglement wedge of $R$ where $\gamma_R(A)$ might be force to enter and leave the entanglement wedge of $R$ multiple times. Whenever a surface has to enter and leave an entanglement wedge, the constraint is lifted. So, if we had two connected components of the entanglement wedge $r_1,r_2$ and $\gamma_R(A)$ had to exit $r_1$ as well as enter and exit $r_2$, we would only constrain it as $\gamma_{R_1}(A)$. 

\section{Further examples and causal relations}

When the HRT surfaces $\gamma(R),\gamma(A)$ are not in the same Cauchy slice (i.e. $\bar{S}_R(A) \not = S(A)$), $\gamma(R)$ may intersect with the future domain of influence of $\gamma(A)$, or the past domain of influence, or both. In two boundary dimensions when $\gamma(R)$ and $\gamma(A)$ each have a single component, it seems that there is a well-defined causal ordering between these two surfaces. In other words, 
there exist $\gamma (A) \in \Sigma_A,\gamma (R) \in \Sigma_R$ such that $\Sigma_R$ is either entirely in the future or entirely in the past of $\Sigma_A$:  $\Sigma_R \in J^{\pm}[\Sigma_A]$.
We will leave the rigorous proof for future works, and focus on the case when such a causal ordering between $\gamma(R)$ and $\gamma(A)$ can be defined. One could wonder whether this causal structure has any implication for $\gamma_{R}(A)$.  Since $\gamma_{R}(A)$ ends in $\partial A$ (as $\gamma(A)$) and has to intersect $\gamma(R)$, we expect the previous ordering to be preserved: $\Sigma_{R} \in J^{+}[\Sigma_A]$ implies that  there exists some surface $\gamma_R (A)\in \Sigma_{A|R}$ such that $\Sigma_{A|R} \in J^{+}[\Sigma_{A}]$ and similar when it is in the past.

Does this ordering have some boundary interpretation? We would like to conjecture that, for the case of constant boost angle this is related with the sign of the minima of modular flow:
\begin{equation}
\Sigma_R \in \left\{ \begin{array}{cc}J^{+}[\Sigma_{A}],&\text{if~}s_{\min}>0\\
J^{-}[\Sigma_{A}],&\text{if~}s_{\min}<0\end{array}\right.
\end{equation}

It is easy to see how this works for the thermofield double, and other cases with local modular flow. See figure \ref{fig:localTFD} for an illustration: for positive $t$, we have $\gamma(R)\in J^-[\gamma(A)]$, and we have $s_{min}<0$ on the boundary. In $d=2$, where we have the argument in terms of heavy twist operators, one can also understand how this works. As was shown in \cite{Faulkner:2018faa},  $s_{min}$ should be interpreted as the relative local boost between the two geodesics as they intersect $\gamma(R)$. In this way, their relative boost determines whether these geodesics are pointing towards the future or the past respectively. Once this causal relation is chosen locally, this fixes the global causal structure, as long as there is a causal relation between $\gamma(A),\gamma(R)$. Of course, if there is no global causal relation between these two surfaces, the sign of $s_{min}$ will only give the local causal structure. We conjecture that this is also true in higher dimensions: whenever there is the causal relation between the surfaces, the sign of $s_{min}$ determines whether it is in the past or in the future, even if $\bar{S}_R(A) \not =\frac{|\gamma_R(A)|}{4 G_N}$ in the absence of constant-boost constrained surfaces. In the case where the is a constant boost surface, we expect that $s_{min}$ is the value of the local boost. 

 In this section we will explore the modular minimal entropy in two example systems: a bulk calculation in the Vaidya geometry and a boundary calculation in a free fermion model.

\subsection{Vaidya geometry}

An interesting example of time-dependent spacetime is matter collapsing and forming a black hole. In holographic theories, this process is dual to the thermalization process in the boundary CFT (see \cite{Hubeny:2007xt,AbajoArrastia:2010yt,Hartman:2013mia,Liu:2013qca} for some references). In this section, we will investigate the collapse of a spherically symmetric, infinitely thin shell of massless particles in 2+1 dimensional AdS spacetime, creating a BTZ black hole. We will focus on the behavior of the constrained surfaces in this geometry, which are then dual to the modular minimal entropies in the thermalization process by our proposal. The metric of the spacetime is given by
\begin{equation}
ds^2 = -f(v,r)dv^2 + 2dvdr +r^2 d\phi^2,
\end{equation}
where $v$ is the ingoing time.
\begin{equation}
f(v,r)  = r^2 + 1 - \theta(v)(r_{+}^2 + 1),
\end{equation}
with $\theta(v)$ the Heaviside step function. For convenience, we have set the AdS radius of curvature $L_{\textrm{AdS}} = 1$. The infinitely thin shell locates at $v=0$. Inside the shell ($v<0$), the spacetime is pure AdS$_3$, while outside the shell ($v>0$), the spacetime is given by the BTZ black hole metric, with event horizon of radius $r_+$. Inside the shell, the static time $t$ is given in terms of $v$ and $r$ by
\begin{equation}
t = v-\tan^{-1}r + \frac{\pi}{2}.
\end{equation}
Let $r\rightarrow\infty$, we find $t_{\infty} = v$ being the boundary field theory time coordinate, and the thin shell starts to fall in at $t=0$.

The geodesic equations for spacelike geodesics are:
\begin{equation}
L=r^2\dot{\phi},
\end{equation}
\begin{equation}
E=f(v,r)\dot{v} - \dot{r},
\end{equation}
\begin{equation}
\dot{r}^2 = E^2 - \left(\frac{L^2}{r^2}-1\right)f(v,r).
\end{equation}
The geodesics whose endpoints lie at equal time on the boundary are studied in \cite{Hubeny:2013dea,Ziogas:2015aja} in detail. However, in general, the constrained surfaces are not constituted of geodesics with endpoints at equal times. Suppose we are looking at the constrained surface of boundary region $A$, denoted by $\gamma_{R}(A)$, that is constrained to cross the HRT surface $\gamma(R)$ of boundary region $R$ once. $\gamma_{R}(A)$ will be the union of two pieces of geodesics $\gamma_{R}(A)_{L}$ and $\gamma_{R}(A)_{R}$, joined on $\gamma(R)$, with extremal total length. 

To do this calculation, we first pick a point $P$ on $\gamma(R)$, then find the geodesics $\gamma_{R}(A)_{L}$ and $\gamma_{R}(A)_{R}$ connecting $P$ to the boundary points of region $A$, then do the minimization of the total length with respect to the position of $P$. In the examples, we choose the horizon radius $r_{+}$ to be equal to the AdS radius. When we calculate the length of the HRT surface or the constrained surface, we need to subtract the divergent part $2\log 2r_{\infty}$. In the following, we will show two examples, one with $|A|<|R|$, and the other with $|A|>|R|$. For the special case of $|A| = |R|$, since there is a spatial $Z_2$ symmetry interchanging $A$ and $R$, the HRT surfaces $\gamma(A)$ and $\gamma(R)$ always cross, so that the constrained surface has the same area as the HRT surface $|\gamma_R(A)| =| \gamma(A)|$.

\begin{figure*}[h!]
    \centering
    \begin{subfigure}[b]{0.3\textwidth}
     \centering
     \includegraphics[height=2in]{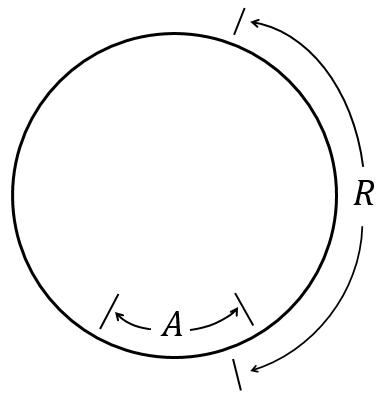}
        \caption{}
        \label{fig:Vaidyashape1}
    \end{subfigure}%
    ~ 
    \begin{subfigure}[b]{0.7\textwidth}
        \centering
        \includegraphics[height=2in]{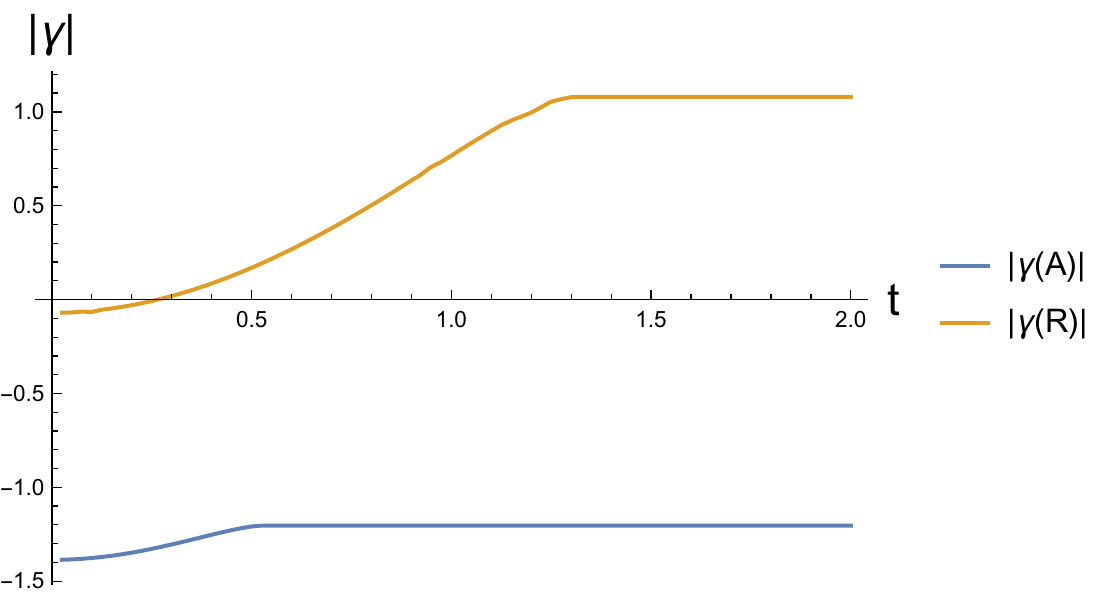}
        \caption{}
        \label{fig:VaidyaHRT1}
    \end{subfigure}
    \caption{(a) Choice of the regions $A$  and $R$. (b) Lengths of the HRT surfaces $|\gamma(A)|$ and $|\gamma(R)|$ as functions of time.}
\end{figure*}

\begin{figure}[h!]
\centering
\includegraphics[width=4in]{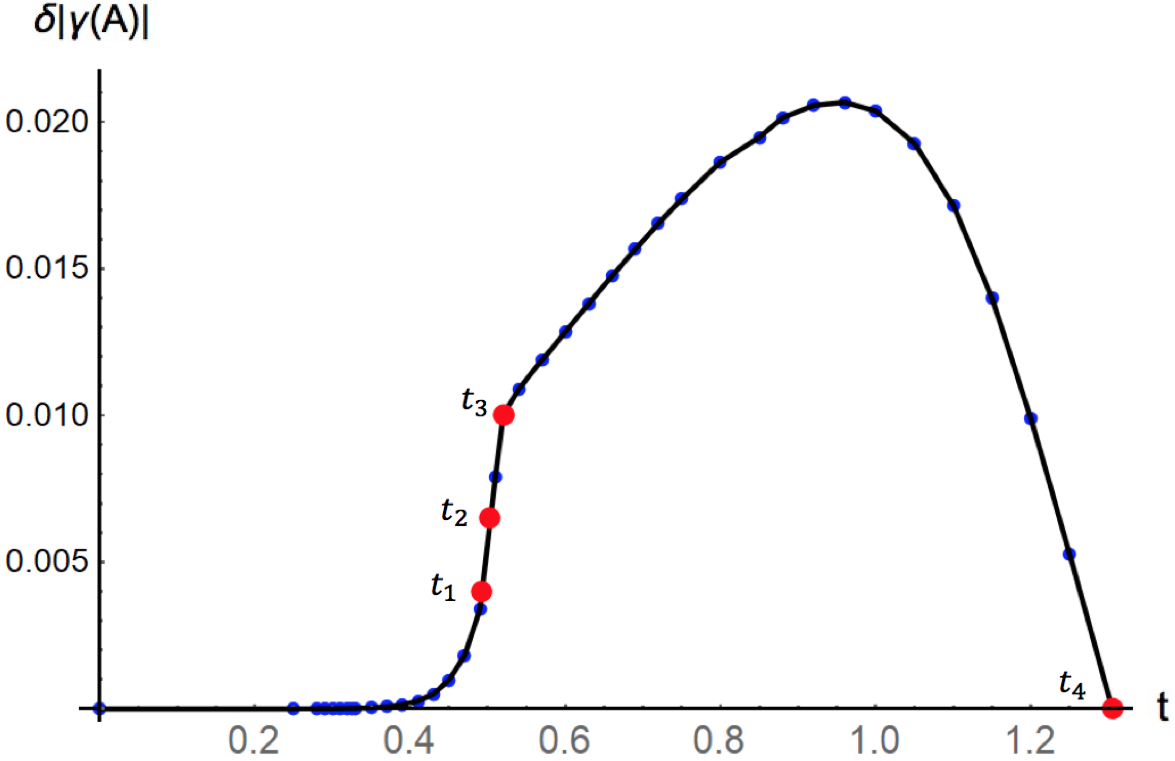}
\caption{$\delta |\gamma(A)| = |\gamma(A)| - |\gamma_{R}(A)|$ as a function of time.}
\label{fig:deltagamma1}
\end{figure}

\noindent{\bf 1. The $|A|<|R|$ case. } In this example, we choose $|A| = \pi/3$, $|R| = 5\pi/6$ and $|A\cap R| = \pi/12$ (as shown in fig. \ref{fig:Vaidyashape1}). In fig. \ref{fig:VaidyaHRT1}, we also calculated the geodesic lengths of HRT surfaces $|\gamma(A)|$ and $|\gamma(R)|$, which grows as functions of $t$. According to the HRT proposal, this growth captures the growth of the entanglement entropies in boundary field theory in the thermalization process. Note that the entanglement entropy of larger region saturates later. In the bulk, the saturation happens when the HRT surface no longer crosses the shell and lies entirely in the BTZ part. Thus, when the entanglement entropies of both regions saturate, the two HRT surfaces will both lie in the static BTZ part of the spacetime, and thus cross each other, in which case the constrained surface coincides with the HRT surface. Before this time, the constrained surface is different from the HRT surface, and their lengths have a finite difference $\delta |\gamma (A)|$. The difference is computed and plotted in fig. \ref{fig:deltagamma1}.

In fig. \ref{fig:deltagamma1}, there are four special points marked by $t_{1,2,3,4}$, which can help us understand the behavior of the various surfaces. When $t<t_1$, the three surfaces $\gamma(A)$, $\gamma_R(A)$, $\gamma(R)$ all cross the infalling shell $v=0$. After time $t_1$, the crossing point $P$ moves outside the shell. After $t_2$ the constrained surface $\gamma_R(A)$ starts to lie entirely in the BTZ part, while the HRT surface $\gamma(A)$ still crosses the shell until $t_{3}$. $t_4$ is the time after which $\gamma(R)$ does not cross the shell anymore, and thus $\delta |\gamma(A)|=0$. 

\begin{figure*}[h!]
    \centering
    \begin{subfigure}[b]{0.5\textwidth}
     \centering
     \includegraphics[height=3in]{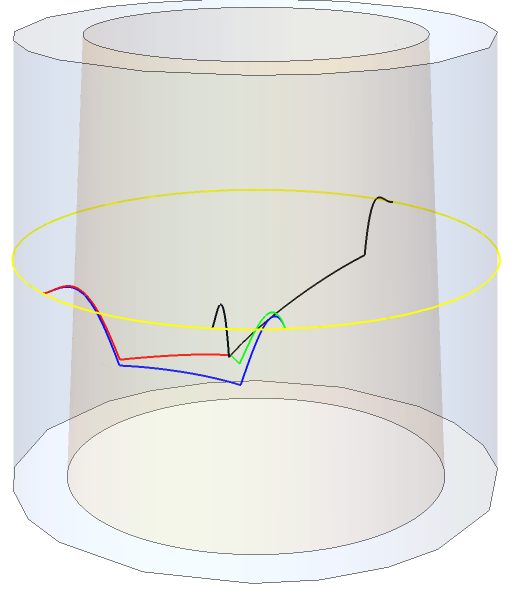}
        \caption{$t=0.4$}
    \end{subfigure}%
    ~ 
    \begin{subfigure}[b]{0.5\textwidth}
        \centering
        \includegraphics[height=3in]{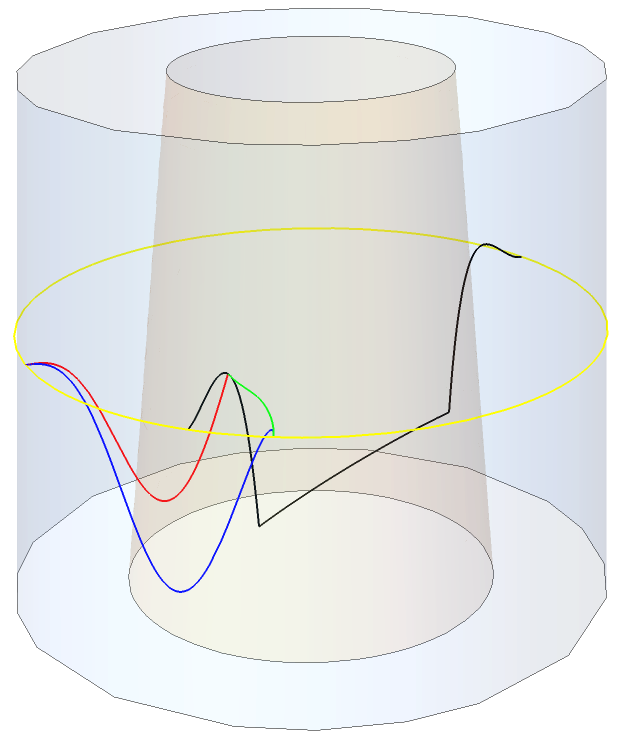}
        \caption{$t=0.7$}
    \end{subfigure}
    \caption{In the figures, the blue cylinder represents the asymptotic boundary, the orange cone denotes the infalling shell and the yellow circle denotes the constant time slice of the boundary. The HRT surfaces $\gamma(A)$ and $\gamma(R)$ are in color blue and black. The constrained surface $\gamma_{R}(A)$ is the union of the red curve and the green curve. }
    \label{fig:Vaidya1}
\end{figure*}

To get a sense of what the constrained surface $\gamma_R(A)$ looks like, in fig. \ref{fig:Vaidya1}, we provide two examples corresponding to two different times. In the left figure, all three surfaces cross the shell, while in the right figure, only the HRT surface $\gamma(R)$ crosses the shell. 

One can intuitively see from fig. \ref{fig:Vaidya1} that in both cases, time orderings can be defined among the surfaces, as the HRT surface $\gamma(A)$ lies in the past of both the HRT surface $\gamma(R)$ and the constrained surface $\gamma_R(A)$.  

In the Vaidya spacetime example, if $|A|<|R|$, we have $\gamma(R)\in J^+[\gamma(A)]$ and $\gamma_R(A)\in J^+[\gamma(A)]$. We will compare this property with the sign of the minima of modular evolution in the field theory example.

\begin{figure*}[h!]
    \centering
    \begin{subfigure}[b]{0.3\textwidth}
     \centering
     \includegraphics[height=1.8in]{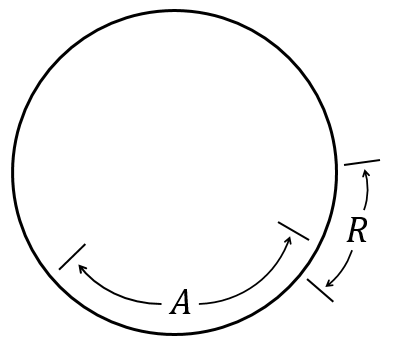}
        \caption{}
        \label{fig:Vaidyashape2}
    \end{subfigure}%
    ~ 
    \begin{subfigure}[b]{0.7\textwidth}
        \centering
        \includegraphics[height=2in]{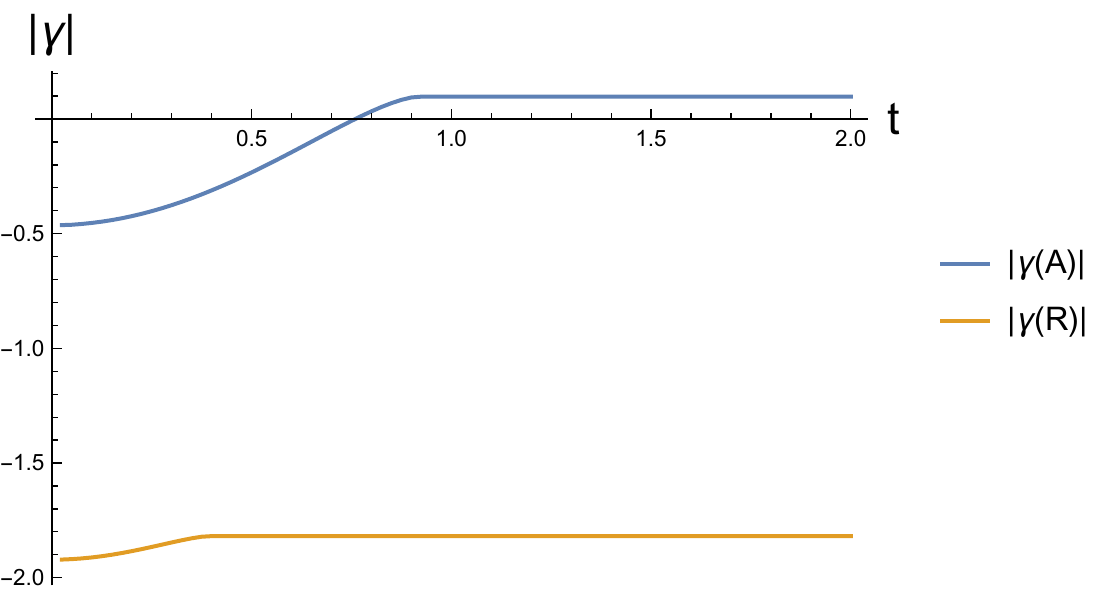}
        \caption{}
        \label{fig:VaidyaHRT2}
    \end{subfigure}
    \caption{(a) Choice of the regions $A$  and $R$. (b) Lengths of the HRT surfaces $|\gamma(A)|$ and $|\gamma(R)|$ as functions of time.}
\end{figure*}

\noindent{\bf 2. The $|A|>|R|$ case.} In this example, we choose $|A| = 7\pi/12$, $|R| = \pi/4$ and $|A\cap R| = \pi/12$ (as shown in fig. \ref{fig:Vaidyashape2}). In fig. \ref{fig:VaidyaHRT2}, we also calculated the geodesic lengths of HRT surfaces $|\gamma(A)|$ and $|\gamma(R)|$. In this case, the length of $\gamma(A)$ saturates later than $\gamma(R)$.

\begin{figure}[h!]
\centering
\includegraphics[width=4in]{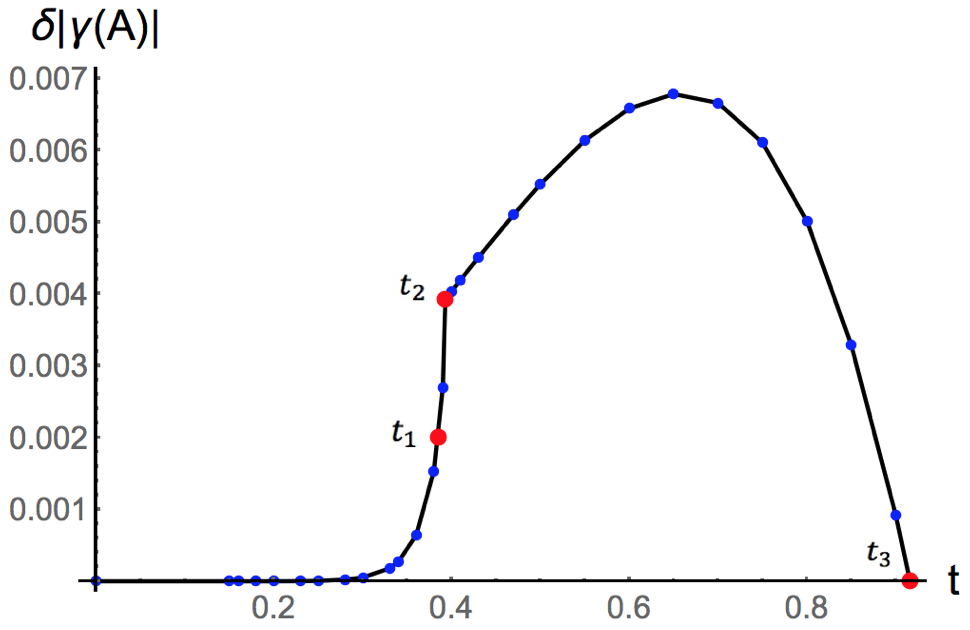}
\caption{$\delta |\gamma(A)| = |\gamma(A)| - |\gamma_{R}(A)|$ as a function of time.}
\label{fig:deltagamma2}
\end{figure}

We computed the difference between $|\gamma_R(A)|$ and $|\gamma(A)|$, as shown in fig. \ref{fig:deltagamma2}. The dependence on time has similar form as the previous case. In the plot, there are three special points marked by $t_{1,2,3}$, whose meanings are different from the previous example. When $t<t_1$, the three surfaces $\gamma(A)$, $\gamma_R (A)$, $\gamma(R)$ all cross the shell $v=0$. After time $t_1$, the crossing point $P$ moves outside the shell, and after $t_2$ the HRT surface $\gamma(R)$ lies entirely in the $v>0$ region, i.e. the BTZ part. At time $t_{3}$, $\gamma_R(A)$ and $\gamma(A)$ become the same surface, and come out of the shell at the same time.

\begin{figure*}[h!]
    \centering
    \begin{subfigure}[b]{0.5\textwidth}
     \centering
     \includegraphics[height=3in]{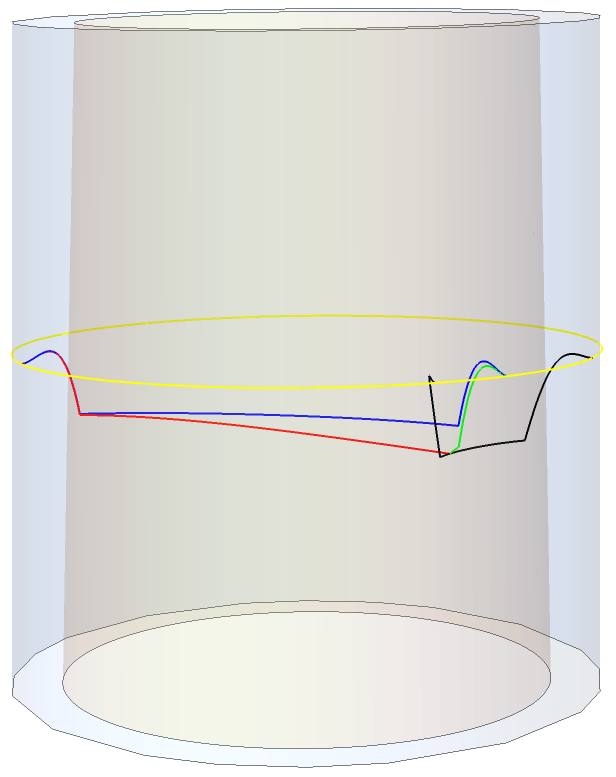}
        \caption{$t=0.4$}
    \end{subfigure}%
    ~ 
    \begin{subfigure}[b]{0.5\textwidth}
        \centering
        \includegraphics[height=3in]{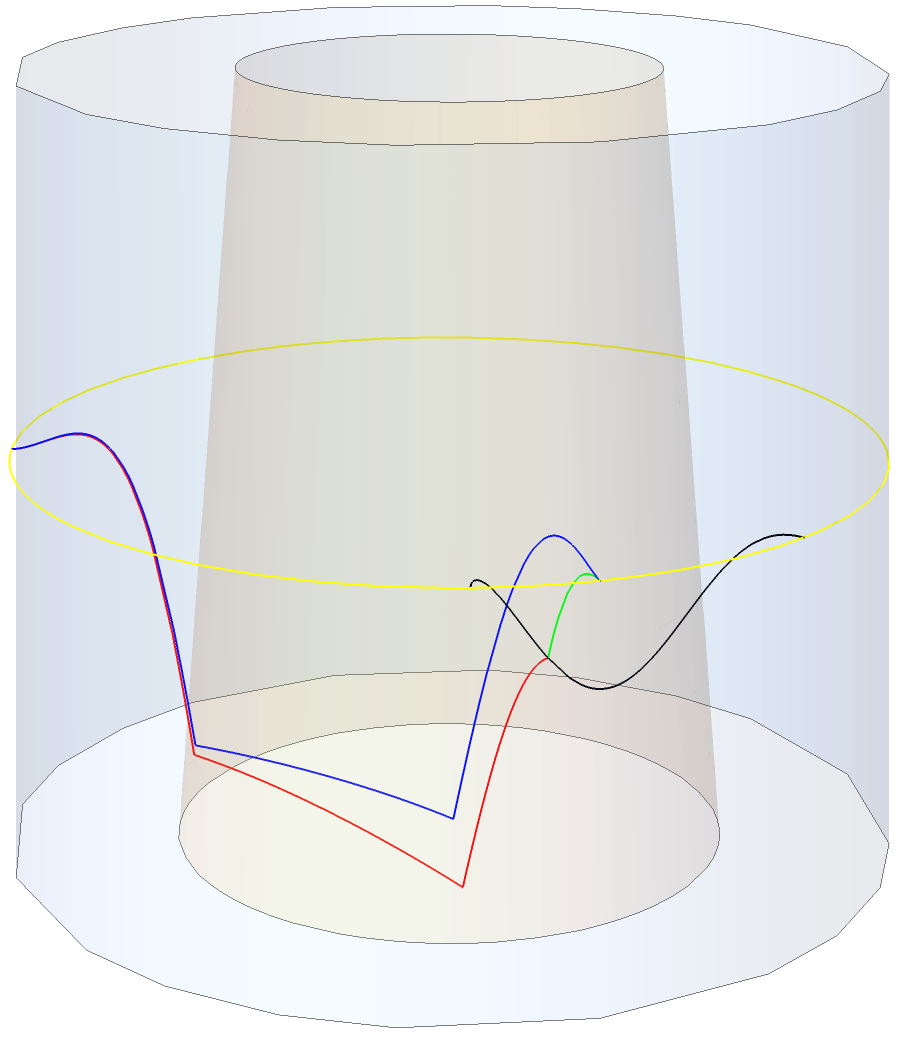}
        \caption{$t=0.7$}
    \end{subfigure}
    \caption{In the figures, the blue cylinder represents the asymptotic boundary, the orange cone denotes the infalling shell and the yellow circle denotes the constant time slice of the boundary. The HRT surfaces $\gamma(A)$ and $\gamma(R)$ are in color blue and black. The constrained surface $\gamma_{R}(A)$ is the union of the red curve and the green curve.}
    \label{fig:Vaidya2}
\end{figure*}

In fig. \ref{fig:Vaidya2}, we provide two examples corresponding to two different times. The left figure corresponds to earlier time when all three surfaces cross the shell. The right figure corresponds to later time when the HRT surface $\gamma(R)$ lies entirely outside the shell. 

As can be seen from fig. \ref{fig:Vaidya2}, in the Vaidya spacetime example, if $|A|>|R|$, we have $\gamma(R)\in J^-[\gamma(A)]$ and $\gamma_R(A)\in J^-[\gamma(A)]$. The time ordering is reversed comparing to the case when $|A|<|R|$.

\subsubsection*{Intuitive explanation}

We believe that the result here is not specific to the particular solution and is a generic feature of chaotic system after a global quench. For a system obeying the eigenstate thermalization hypothesis (ETH) \cite{PhysRevA.43.2046,PhysRevE.50.888}, we can understand this result based on the following argument. For small enough  $R$, ETH tells that $\rho_R \sim e^{-\beta H}$. On the boundary, if the system is undergoing a thermalization process, a negative modular parameter $s_{min}<0$ corresponds to evolving backward in time, which lowers the entropy. In the bulk, we will see in the Vaidya spacetime example, for $|R|\ll|A|$, we have $\Sigma_R\in J^-[\Sigma_A]$.

\subsection{Free fermions}

In holographic theories, our proposal relates the modular minimal entropy of boundary theory to a geometrical object, the constrained surface in the bulk. For theories without a gravity dual, we do not expect the geometric picture to hold. Nevertheless, it remains an interesting question to study the behavior of modular minimal entropy.
To gain some insight beyond holographic theories, in this section, we study the modular minimal entropy in a simple model of free fermions on a lattice.

\begin{figure}[h!]
\centering
\includegraphics[width=3in]{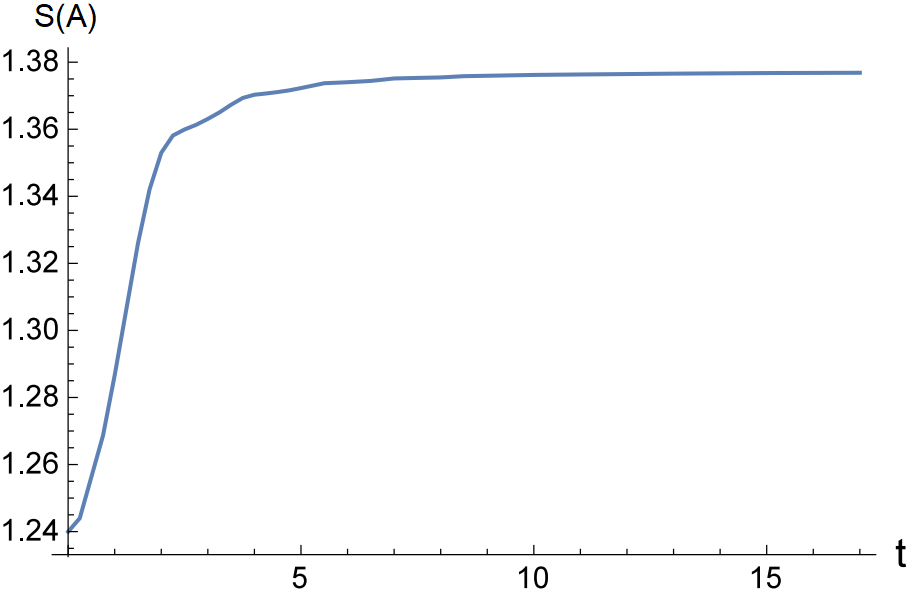}
\caption{The entanglement entropy of a region grows with time and approaches an equilibrium. In the plot, the size of the total system is $L=202$, and the size of the subsystem is $|A|=6$. Fermion mass $m = 1/100$.}
\label{fig:FermionEE}
\end{figure}

We put the fermions on a one-dimensional lattice of length $L$, with the end of the lattice attached to the beginning and forms a loop. The Hamiltonian of the free fermion lattice system is written in a tight binding form
\begin{equation}
H=-\sum_{\langle ij\rangle}c_{i}^{\dagger}c_j + m\sum_{i}(-1)^{i}n_{i},
\end{equation}
where $n_{i}=c_{i}^{\dagger}c_{i}$. The Hamiltonian contains a hopping term (with hopping constant set as one) between nearest neighbor sites and a staggered potential. The staggered potential opens up a gap in the band, and gives the fermions a mass. We first prepare the system in ground state of non-zero mass $m$, and at half filling.  
The Hamiltonian can be diagonalized in momentum space as $H=\sum_k\left(E_{k+}d_{k+}^\dagger d_{k+}+E_{k-}d_{k-}^\dagger d_{k-}\right)$ with $E_{k\pm}=\pm\sqrt{m^2+4\cos^2\frac k2}$. $d_{k\pm}$ are superpositions of $c_i$ that annihilate fermions in the upper and lower bands, respectively. The ground state is 
\begin{equation}
\ket{\psi} = \ket{GS} \equiv \prod_{k}d_{k-}^{\dagger}\ket{vac}.
\end{equation}
At time $t=0$, we quench the system by switching off the fermion mass $m$. The evolution of the system is reflected in the evolution of the operators $d_{k-}^{\dagger}(t)$, i.e.,
\begin{equation}\label{freefermion}
\ket{\psi(t)} = \prod_{k}d_{k-}^{\dagger}(t)\ket{vac}.
\end{equation}
For states having the form of (\ref{freefermion}) that Wick theorem applies, the reduced density matrix of a subsystem can be calculated through correlation functions (see for example \cite{2003JPhA...36L.205P}). The reduced density matrix of region $R$ can be written as
\begin{equation}
\rho_{R}(t) = \frac{1}{Z_R} \exp\left(- \sum_{i,j\in R}H_{R,ij}c_{i}^{\dagger}c_{j}\right),
\end{equation}
where $Z_{R}$ is a normalization factor, and matrix $H_{R}$ is determined by
\begin{equation}
H_{R}^{T}=\log \left(\frac{1-C}{C}\right).
\end{equation}
In the equation, $T$ means transpose and $C$ is the correlation matrix defined as
\begin{equation}
C_{ij} = \bra{\psi(t)} c_{i}^{\dagger} c_{j}\ket{\psi(t)}.
\end{equation}
We can further use the reduced density matrix to calculate the entanglement entropy $S(R)$. After the quench, the entanglement entropy $S(A)$ of region $A$ will increase and then saturate at a maximum value\footnote{It should be noted that the free fermion system is integrable, such that the entropy saturation only lasts for a short time proportional to $L-|A|$, in contrast to thermalizing system (such as the dual of Vaidya geometry we studied in previous subsection), which stays in equilibrium for exponentially long time.} (see fig. \ref{fig:FermionEE}). 

Due to the quadratic form of the modular Hamiltonian, if we use $\rho_{R}(t)$ to do a modular flow on the system, 
\begin{equation}
\rho (t,s) = \rho_{R}^{is}\ket{\psi(t)}\bra{\psi(t)} \rho_{R}^{-is}=\ket{\psi(t,s)}\bra{\psi(t,s)},
\end{equation}
the form of the state $\ket{\psi(t,s)}$ is preserved as
\begin{equation}
\ket{\psi(t,s)} = \prod_{k}d_{k-}^{\dagger}(t,s)\ket{vac}.
\end{equation}
Thus the above method of calculating reduced density matrix still applies. 

\begin{figure*}[h!]
    \centering
    \begin{subfigure}[b]{0.5\textwidth}
     \centering
     \includegraphics[height=2in]{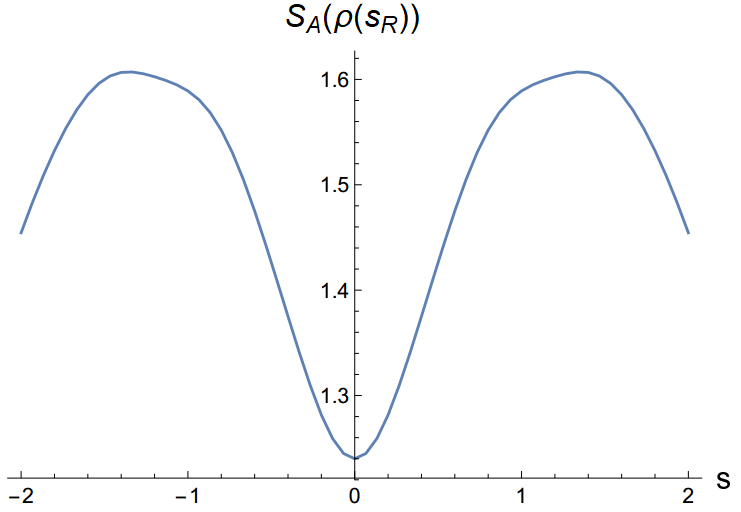}
        \caption{$t=0$}
    \end{subfigure}%
    ~ 
    \begin{subfigure}[b]{0.5\textwidth}
        \centering
        \includegraphics[height=2in]{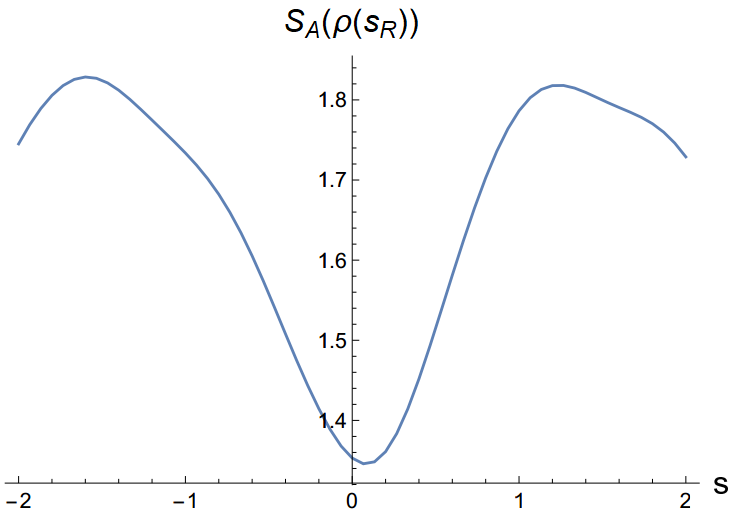}
        \caption{$t=2$}
    \end{subfigure}
    \caption{The entanglement entropy of region $A$ as a function of the modular parameter $s_{R}$. (a) $t=0$ (before the quench). (b) $t=2$ (after the quench). }
    \label{fig:FermionMF}
\end{figure*}

Before we study how the modular minimal entropy behaves after the quench, we can first look at how the modular flow changes the entanglement entropy in the initial state. In the numerics, We choose $m=1/10$, fix the total size of the system being $L=202$, and choose regions $A$ and $R$ with $|A|=6$, $|R|=10$, $|A\cap R|=3$. The entanglement entropy of region $A$ as a function of the modular parameter $s_{R}$ is shown in fig. \ref{fig:FermionMF}.(a). From the figure, we can see that $s=0$ corresponds to a local minimum of the entanglement entropy, which is expected since the ground state preserves time reversal symmetry. After the quench, the local minimum position $s_{min}$ will be shifted away from zero, as shown in fig. \ref{fig:FermionMF}.(b). 

\begin{figure*}[h!]
    \centering
    \begin{subfigure}[b]{0.5\textwidth}
     \centering
     \includegraphics[height=2in]{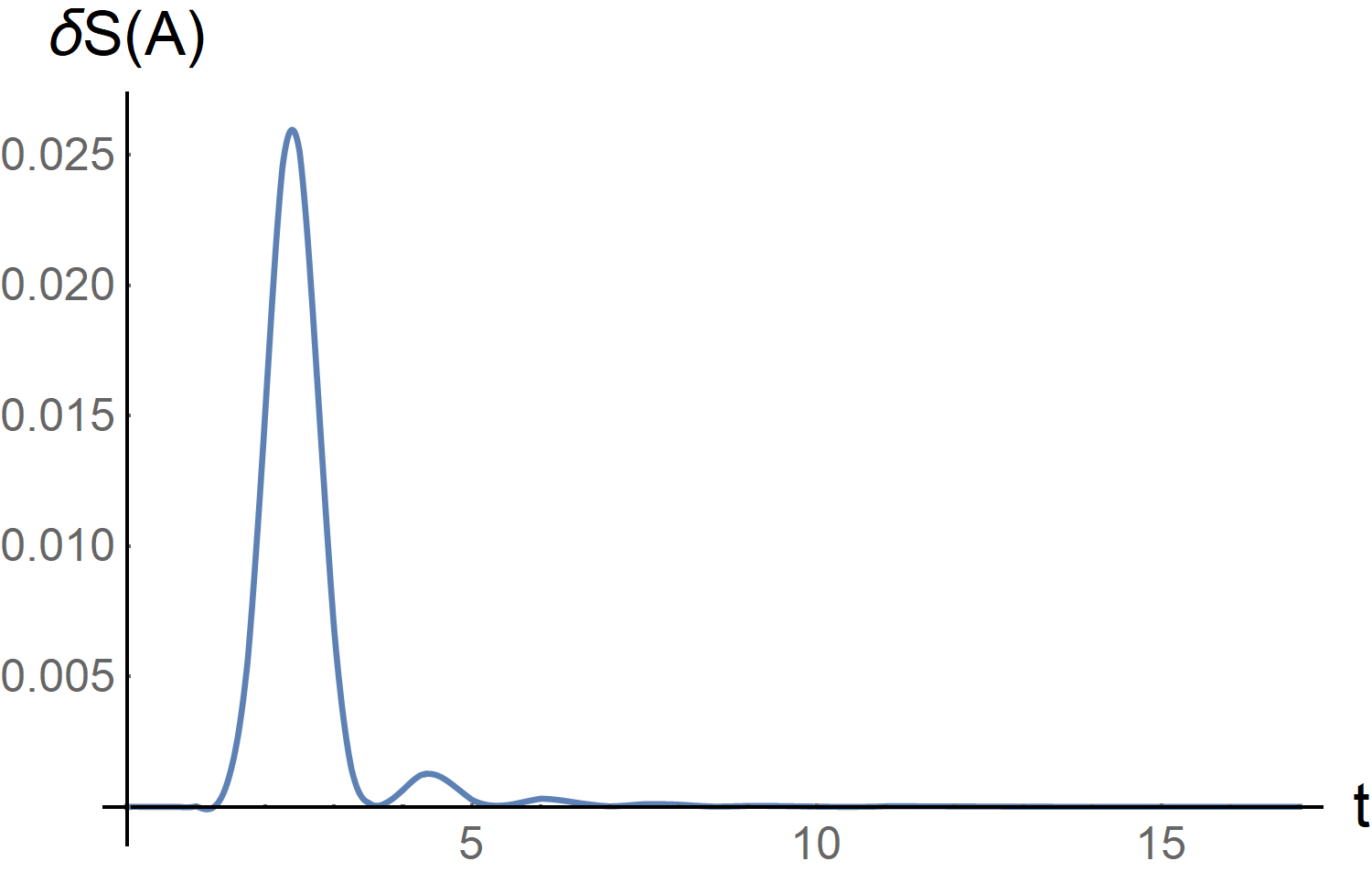}
        \caption{$\delta S(A)-t$}
    \end{subfigure}%
    ~ 
    \begin{subfigure}[b]{0.5\textwidth}
        \centering
        \includegraphics[height=2in]{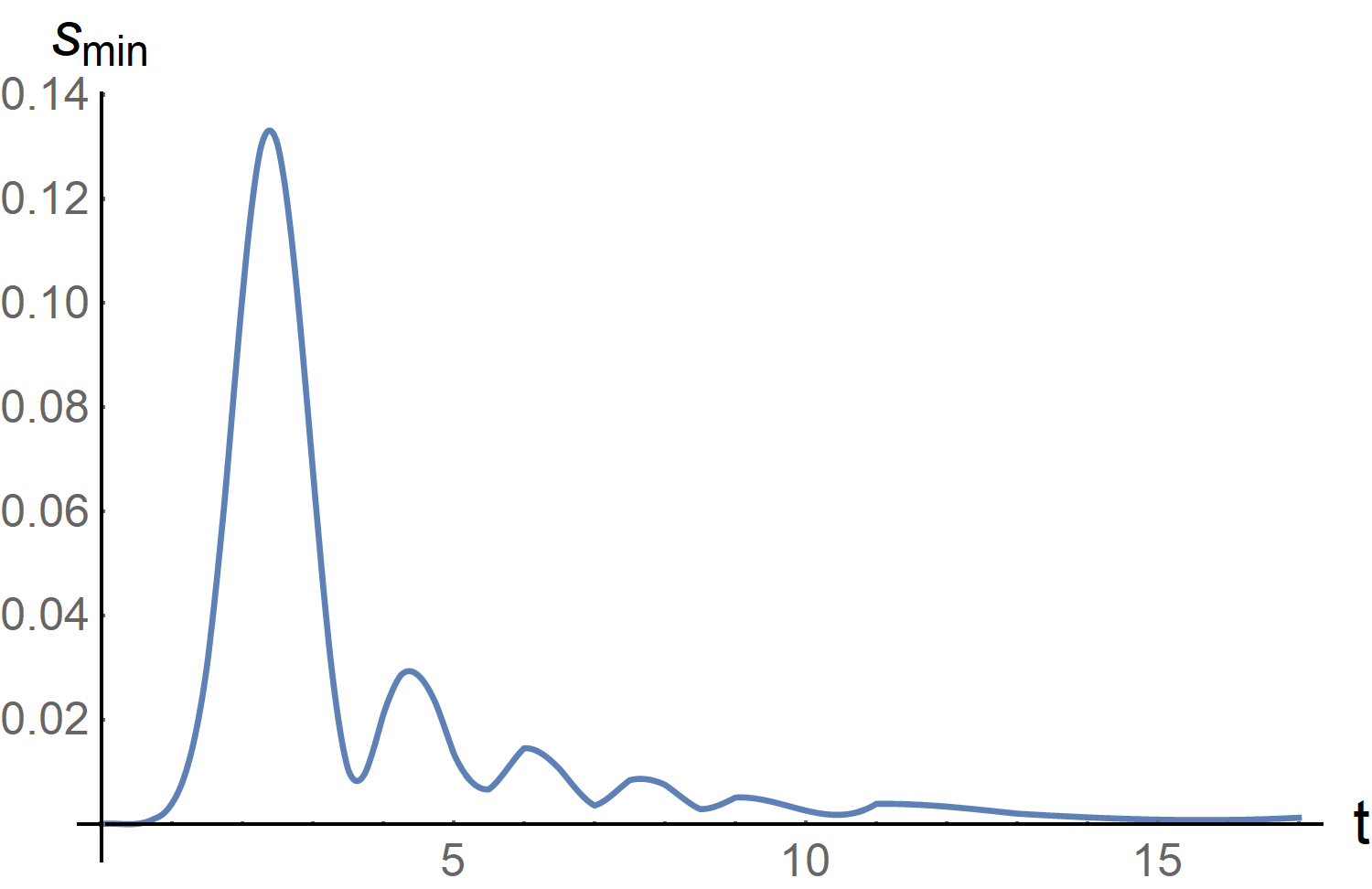}
        \caption{$s_{min}-t$}
    \end{subfigure}
    \caption{In the plots, $m=1/10$, $L=202$, and $|A|=6$, $|R|=10$, $|A\cap R|=3$. (a) $\delta S(A)-t$ curve. (b) $s_{min}-t$ curve. }
    \label{fig:Fermionquench}
\end{figure*}

After the quench, the modular minimal entropy of region $A$ will grow, similar to the behavior of the entanglement entropy. We focus on the difference $\delta S(A) = S(A)- S_R(A)$, which, as shown in Fig. \ref{fig:Fermionquench}, increases and then decreases. As entropy reaches the saturation value, $\delta S(A)$ returns to zero, which is qualitatively the same as the Vaidya geometry case. Another quantity that we are interested in is the modular parameter $s$ at the minimum, which we denote as $s_{\min}$. When $s_{\min}$ is small, by taking a quadratic approximation around the $s=0$ curve we see that $\delta S(A)\propto s_{min}^2$ at lowest order. $s_{min}$ as a function of $t$ is shown in fig. \ref{fig:Fermionquench}.(b). The oscillations in the figure is a result of the finite bandwidth (UV cutoff) of the system. For small perturbation $m\ll 1$, we can expand $s_{min}$ in terms of $m$. since $s_{\min}$ is odd under the $Z_2$ transformation $m \leftrightarrow -m$, we have $s_{min}\propto m^2$, and thus $\delta S(A) \propto m^4$.

\begin{figure*}[h!]
    \centering
    \begin{subfigure}[b]{0.3\textwidth}
     \centering
     \includegraphics[height=1.3in]{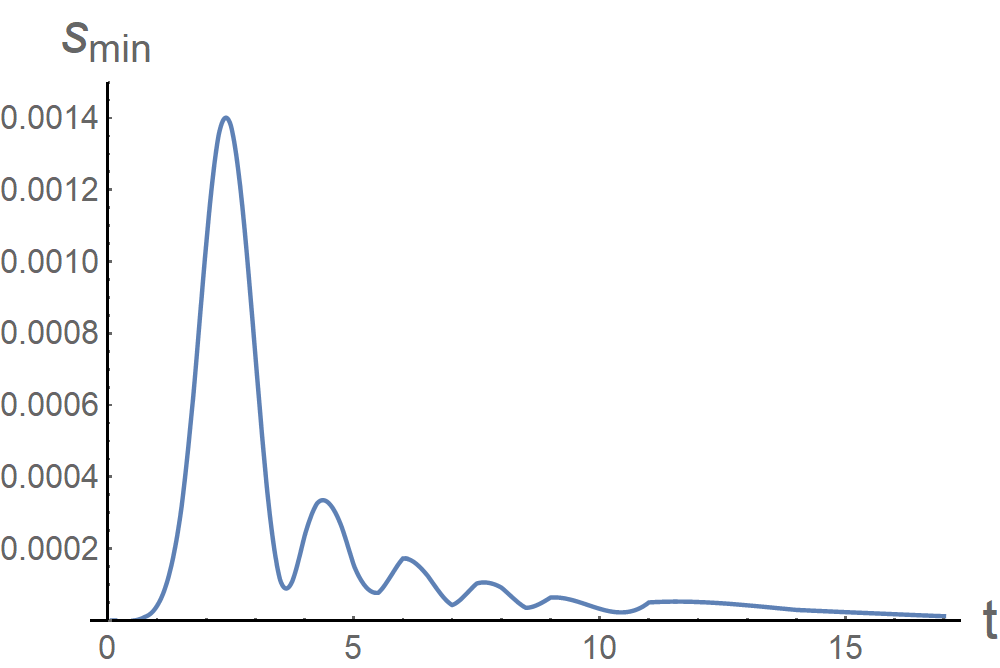}
        \caption{}
    \end{subfigure}%
    ~ 
        \begin{subfigure}[b]{0.3\textwidth}
     \centering
     \includegraphics[height=1.3in]{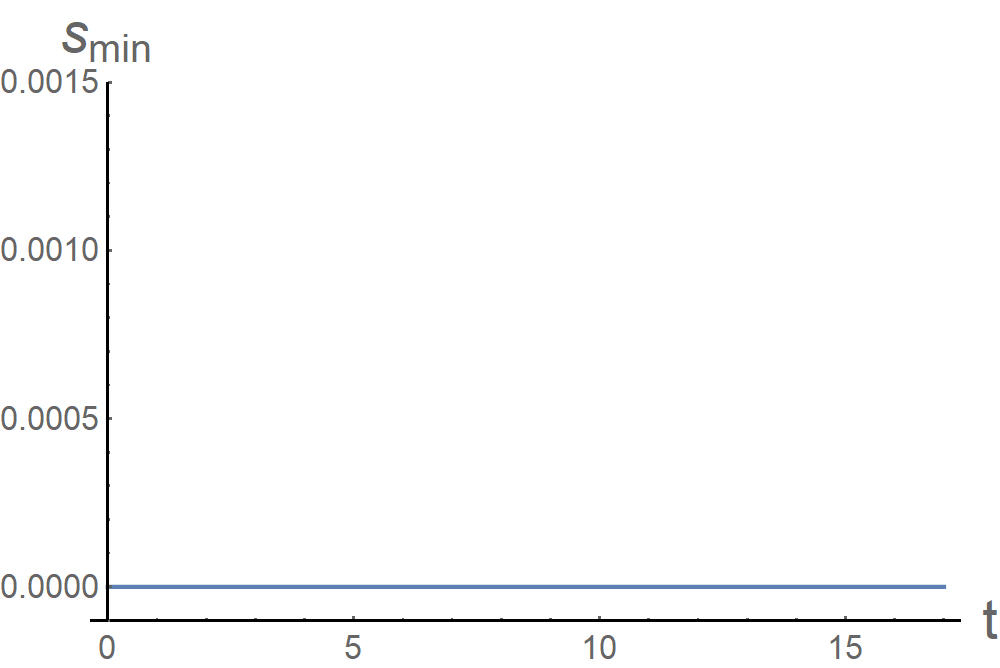}
        \caption{}
    \end{subfigure}%
    ~ 
    \begin{subfigure}[b]{0.3\textwidth}
        \centering
        \includegraphics[height=1.3in]{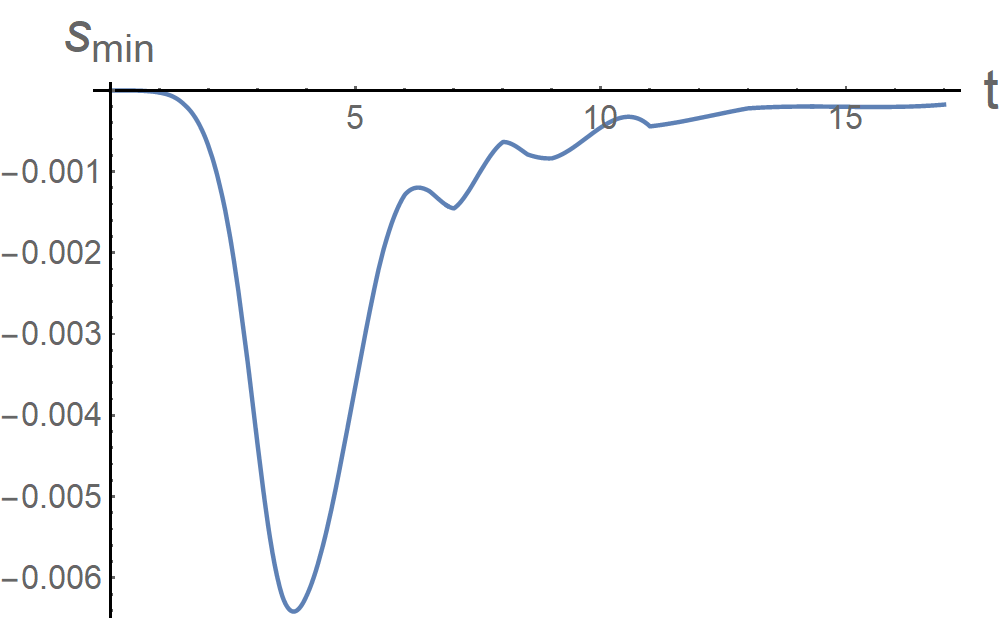}
        \caption{}
    \end{subfigure}
    \caption{(a) $|A| =6, |R|=10, |A\cap R|=3$, (b) $|A|=|R|=2|A\cap R|=10$, (c) $|A|=18, |R|=10, |A\cap R| = 9$. $m = 1/100$ and $L = 202$ are chosen. As the figure shows, $s_{min}$ is positive when $|A|<|R|$, negative when $|A|>|R|$, and stays zero when $|A|=|R|$.}
    \label{fig:sminsign}
\end{figure*}

In the above example, $|A|<|R|$, and we observe that $s_{min}>0$. In fig. \ref{fig:sminsign}, we show that the sign of $s_{\min}$ changes when the order of $|A|$ and $|R|$ switches, just like the Vaidya case. When $|A|=|R|$, $s_{min}$ stays at zero as expected.

\section{Discussion}
\subsection{Modular flow with multiple nested regions}

\begin{figure*}[h!]
\centering
     \includegraphics[height=2.7in]{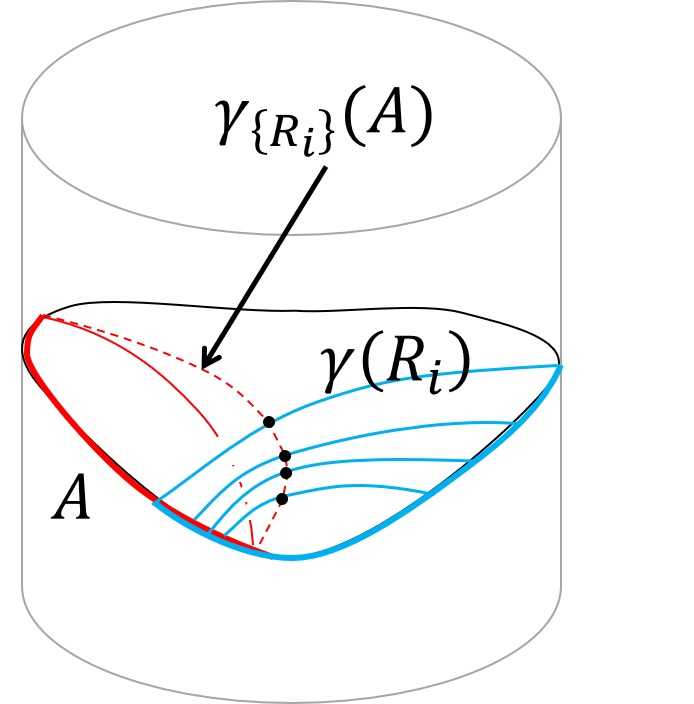}
    \caption{Illustration of a constrained surface (red dashed curve) that is extremal except at intersections with multiple HRT surfaces $\gamma(R_i)$ (blue curves), with nested regions $R_i$.}
    \label{fig:multiregion}
\end{figure*}

As explained, given a region $A$, we generically expect a non-trivial modular minimal entropy with respect to $R$ as long as $R, A$ are partially overlapping. As a generalization, we can constrain $\rho_A$ with respect to a nested set of subregions ${\cal R} \equiv \lbrace R_1, R_2, ... ,R_m \rbrace$, with $R_i \subset R_{i+1}$ (see Fig. \ref{fig:multiregion}). To obtain a nontrivial result, each of the regions should have non-trivial overlap with $A$ and its complement. According to Ref. \cite{Wall:2012uf}, the entanglement wedge of $R_i$ are also nested in the bulk, and, because of this nesting, the bulk constrained surface will remain space-like \footnote{If we considered an arbitrary ordering of the regions, the constrained surface will have kinks and become time-like.}. 
 In this way, we can defined the ${\cal R}$-modular minimal entropy:
\begin{equation}
    \bar{S}_{\cal R}(A)=\text{min}_{s_{\cal R}} S_A(\rho(s_{\cal R}))=\frac{|\gamma^{>}_{\cal R}(A)|}{4 G_N} \ge \frac{|\gamma_{\cal R}(A)|}{4 G_N}
\end{equation}
where
\begin{equation}
    |\psi(s_{\cal R})\rangle \equiv e^{-i K_m s_m} ...e^{-i K_2 s_2} e^{-i K_1 s_1} |\psi \rangle ; ~~~ \gamma_{\cal R}(A)=\text{min}_{\gamma(R_m)...\gamma(R_2) \gamma(R_1)} \gamma, ~~ \partial \gamma=\partial A
\end{equation}
and $\gamma_{\cal R}^>(A)$ is the constrained surface at constant boost angle.
In this case $s_{\cal R}=(s_1,...s_m)$ is a vector whose values are such that they minimize the entropy. There are many directions in which one could explore this ${\cal R}$-modular minimal entropy, such as taking the continuum limit or understand other orderings, which we leave for future work. In the next subsection, we will explore the case of two regions ${\cal R}=\lbrace R_1,R_2 \rbrace$, where there seem to be some interesting inequalities. 

\subsection{Strong subadditivity-like inequalities for 2-modular minimal entropies}

An interesting direction to explore this further is to consider modular evolution with respect to two regions $R_1,R_2$. In this situation, we will have $4$ bulk surfaces of interest: $\gamma(A),\gamma_1(A),\gamma_2(A),\gamma_{12}(A)$ (see figures \ref{fig:exampleinter}, \ref{fig:exampleinter2}), for simplicity of notation we will keep track of the constraining surface by a subindex and $12$ means that it is constrained to go through both $R_1$ and $R_2$. 

Here, we would like to understand whether there is some relation between these four surfaces. As has been discussed,  we expect that $\gamma(R_1),\gamma(R_2)$ will be past/future related with $\gamma(A)$: either $\gamma(R_1),\gamma(R_2) \in J^{\pm}[\gamma(A)]$ or $\gamma(R_1) \in J^{\pm}[\gamma(A)],\gamma(R_2) \in J^{\mp}[\gamma(A)]$ \footnote{Where, by this causal relation between surfaces, we mean that causal relation between Cauchy surfaces that include them, as in section 4.}. In the boundary these two cases differ by the relative sign of $s_{1,min}, s_{2,min}$. When we do the combined modular flow, we expect the signs of $s_{min}$ to interfere constructively or destructively, depending on whether they are the same or different. Let us analyze these two cases from the bulk point of view, illustrated by figures \ref{fig:exampleinter}, \ref{fig:exampleinter2}, respectively.

\subsubsection*{Constructive interference (same signs of $s_{min}$)}

\begin{figure*}[h!]
    \centering
    \begin{subfigure}[b]{0.5\textwidth}
     \centering
     \includegraphics[height=1.6in]{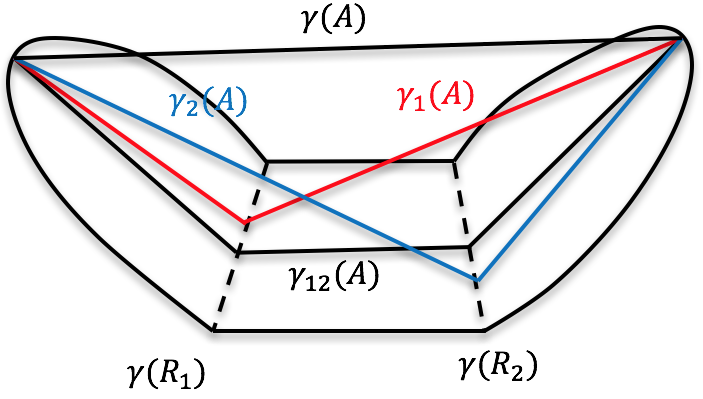}
        \caption{}
    \end{subfigure}%
    ~ 
    \begin{subfigure}[b]{0.5\textwidth}
        \centering
        \includegraphics[height=1.6in]{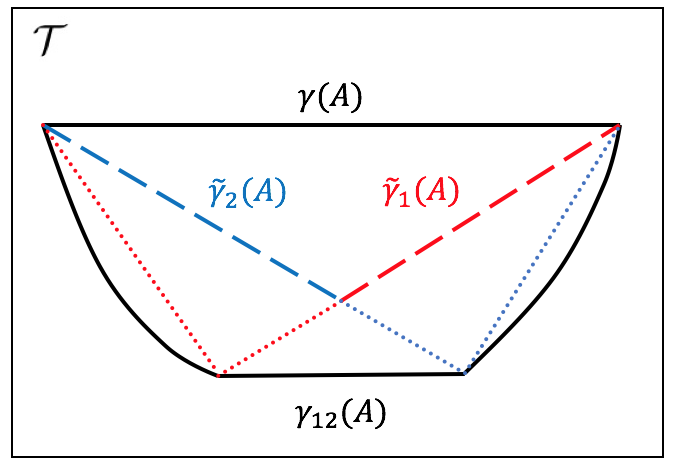}
        \caption{}
    \end{subfigure}
    \caption{This is the case for constructive interference. In the right figure we see how we can apply the strong subadditivity (SSA) proof to this situation by projecting the surfaces $\gamma_1,\gamma_{2}$ to the same slice where $\gamma,\gamma_{12}$ are. We recombine the projected surfaces $\tilde{\gamma}_1$ and $\tilde{\gamma}_2$ into the dashed part and the dotted part, then the length of the dashed part is less than $\gamma$, while the length of the dotted part is less than $\gamma_{12}$. }
    \label{fig:exampleinter}
\end{figure*}

When we have that $\gamma(R_1),\gamma(R_2) \in J^{\pm}[\gamma(A)]$, since we expect constraining to preserve the ordering (see previous section
), we have that $\gamma_1(A),\gamma_2(A),\gamma_{12}(A) \in J^{\pm}[\gamma(A)]$. This implies that we can construct a time-like surface ${\cal T}$ where $\gamma (A),\gamma_{12}(A)$ lie.
We can project $\gamma_1(A)$ and $\gamma_2(A)$ to surface $\cal{T}$ as $\tilde{\gamma}_{1}(A)$ and $\tilde{\gamma}_{2}(A)$, which because of maximin will have larger area: $\gamma_1(A)<\tilde{\gamma}_{1}(A)$ and $\gamma_2(A)<\tilde{\gamma}_{2}(A)$. In this time-like surface where all these four surfaces intersect we can apply the arguments of \cite{Headrick:2007km,Wall:2012uf} (but with an opposite sign since these are maximal surfaces in ${\cal T}$) where we think of $\gamma_{12}(A)+\gamma(A)$ as the sum over the areas of two surfaces which have the same boundary conditions as $\gamma_{1}(A),\gamma_{2}(A)$ (see fig \ref{fig:exampleinter}). We conclude that
\begin{equation}
    \gamma(A)+\gamma_{12}(A)> \tilde{\gamma}_1(A)+\tilde{\gamma}_2(A)>\gamma_{1}(A)+\gamma_{2}(A)
\end{equation}
which we can write in terms of boundary quantities when $\gamma_R=\gamma^{>}_R$:
\begin{equation}
    S(A)-\bar{S}_{R_1}(A)>\bar{S}_{R_2}(A)-\bar{S}_{R_1,R_2}(A) \ge 0
\end{equation}
This expression suggests that by adding further constraints in a constructive way (with the same sign of $s_{min}$), we can reduce the entropy further. 

\subsubsection*{Destructive interference (different signs of $s_{min}$)}

\begin{figure*}[h!]
    \centering
    \begin{subfigure}[b]{0.5\textwidth}
     \centering
     \includegraphics[height=1.6in]{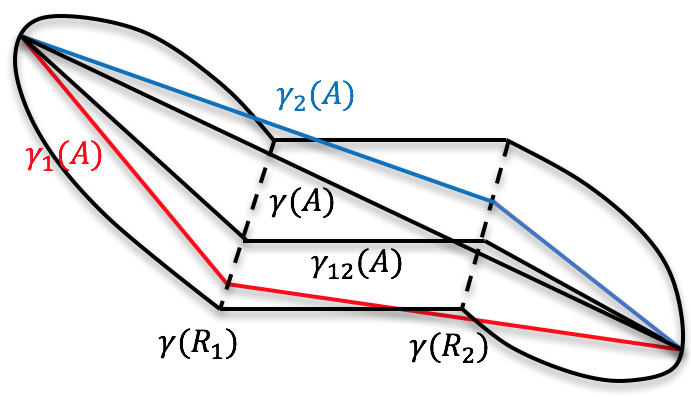}
        \caption{}
    \end{subfigure}%
    ~ 
    \begin{subfigure}[b]{0.5\textwidth}
        \centering
        \includegraphics[height=1.6in]{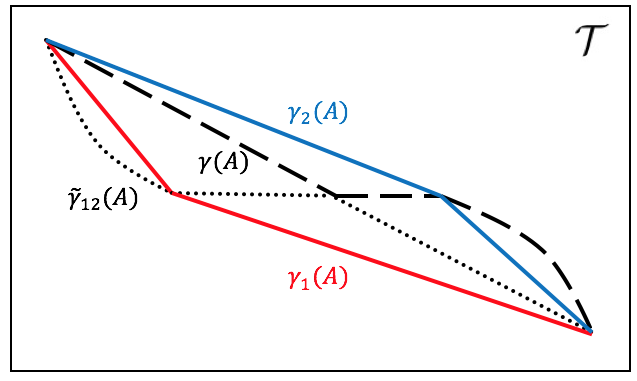}
        \caption{}
    \end{subfigure}
    \caption{This is the case for destructive interference. In (b) we see how we can apply the SSA proof to this situation by projecting the surfaces $\gamma_{12}$ to the same slice where $\gamma_1,\gamma_2, \gamma$ are. We recombine $\gamma$ and the projected surface $\tilde{\gamma}_{12}$ into the dashed part and the dotted part. The length of the dashed part is less than $\gamma_{2}$ and the length of the dotted part is less than $\gamma_1$. }
    \label{fig:exampleinter2}
\end{figure*}

In the opposite case, since $\gamma(R_1),\gamma(R_2)$ are in opposite orderings with respect to $\gamma(A)$, this implies that $\gamma_1(A) \in J^{\pm}(\gamma_2(A))$. Note that, in contrast with the previous case, where $\gamma_{12}(A)$ was in the future or past of $\gamma(A)$ (and thus one could set a time-like surface that interpolates between them), in this case, there is a time-like surface ${\cal T}$ where all $\gamma_{1}(A),\gamma_{2}(A),\gamma(A)$ lie. Therefore, the construction goes in the opposite way: we should project $\gamma_{12}(A)$ to ${\cal T}$, $\tilde{\gamma}_{12}(A)>\gamma_{12}(A)$ and then divide $\gamma(A),\gamma_{12}(A)$ into two surfaces with less area than $\gamma_{1}(A),\gamma_2(A)$ (see \ref{fig:exampleinter2}). We obtain:
\begin{equation}
    \gamma_1(A)+\gamma_2(A)>\gamma(A)+\gamma_{12}(A)
\end{equation}
which we can write in terms of boundary quantities when $\gamma_R=\gamma^{>}_R$:
\begin{equation}
    \bar{S}_{R_2}(A)-\bar{S}_{R_1 R_2}(A)\ge S(A)-\bar{S}_{R_1}(A) \ge 0
\end{equation}
This equation shows that adding the extra constraint on $R_2$ decreases the entropy reduction by constraint on $R_1$.

 At this point it is not clear to us if these inequalities are purely information theoretical or are only true for holographic theories, as for the positivity of tripartite mutual information \cite{Hayden:2011ag}.

\subsection{Relation to the Entanglement of purification}

The entanglement of purification is a function of two overlapping subregions $A, R$. It is defined by minimizing the entropy of $A$ over all states in the Hilbert space that have the same density matrix in $R$:
\begin{equation}
    E_P(R,A)=\text{min}_{\rho_R=\text{tr}_{\bar{R}} |\tilde{\psi} \rangle \langle \tilde{\psi}|} S(\rho_{A,\tilde{\psi}})
\end{equation}

In comparison, our modular minimal entropy is obtained from a constrained minimization, where we only minimize with respect to states generated by the modular flow of $R$ and not all states that preserve the density matrix. Therefore, by definition, it satisfies:
\begin{equation}
    E_P(R,A) \le \bar{S}_{R}(A)
\end{equation}

In \cite{Takayanagi:2017knl,Nguyen:2017yqw}, a holographic proposal for this quantity was put forward \footnote{Note that we are using an equivalent but different notation, more suited to compared it with the modular minimal entropy. In our notation $R_{ours}=A \cup B$, $A_{ours}=A \cup A'$, $R_{ours} \cap A_{ours}= A$.}, in terms of the minimal surface anchored to $\gamma(R)$ which could be in principle extended to end in $\partial A$, but does not intersect with the complement of the entanglement wedge of $R$ (so this surface only ends in the boundary if $\partial A \cap R \not = 0$).  Our bulk construction has some similarities, but in our case, the surfaces are always anchored to the boundary. While one could try to get something similar to the entanglement of purification by averaging our modular minimal entropies over boundary regions, this procedure is not completely satisfactory. In particular, from our perspective it is not clear how one could get rid of the boundary UV divergences using modular flow. 

Based on the conjecture of Ref. \cite{Takayanagi:2017knl,Nguyen:2017yqw} on entanglement of purification and our conjecture on modular minimal entropy, we can geometrically derive the following inequality:
\begin{equation}
    E_P(R,A)+E_P(\bar{R},A) \le \bar{S}_{R}(A)
\end{equation}
since $ E_P(R,A)+E_P(\bar{R},A) $ describes an extremal surface that ends in $\partial A$ and intersects $\gamma(R)$, but the respective contributions from the entanglement wedge of $R,\bar{R}$ don't have to be glued along $\gamma(R)$. Therefore this area can clearly be made lower than the constrained surface, which is connected along $\gamma(R)$. 
\subsection{Higher dimensions}

While we have mainly focused in $d=2$, we believe our conjecture holds in higher dimensions: with a precise equality when there exists a constant boost constrained surface and an inequality when there is none. It would be nice to have a proof of the modular minimal entropy formula in higher dimensions, which would require understanding twist operators in higher dimensions, maybe along the lines of \cite{Hung:2014npa,Dong:2016fnf}.

It would be interesting to explore further how the modular evolved states $|\Psi(s_R) \rangle$ change the divergence structure of $S_A$. We expect that this divergence structure is determined by geometric invariant terms in the codimension-$3$ surface $\partial A \cap \partial R$ and that such terms are subleading to the area law divergence. These divergences were studied for singular space-like corners in \cite{Myers:2012vs}. 

A related aspect that would be nice to understand is to look for a boundary interpretation for the constrained surfaces without a constant local boost angle, and to give a more formal proof of its boundedness by the modular minimal entropy.

\subsection{Quantum corrections}

From the boundary definition of this object, the quantum ($1/N$) corrections are completely well defined. Following \cite{Faulkner:2013ana}, the bulk quantum correction corresponds to the bulk entanglement entropy in the $|\psi(s_{\min})\rangle$ state. This contribution, while well defined, would require understanding the dual to the modular evolved state (as opposed to the entropy of $A$ in this state), maybe using the discussion of \cite{Jafferis:2014lza}. Furthermore, given that bulk and boundary modular flow are the same, one might be able to compute these quantum corrections directly from bulk modular evolution. In other words, the equality between bulk and boundary modular flow seems to naturally imply:
\begin{equation}
    \bar{S}_R(A)=\frac{|\gamma_R^{>}(A)|}{4 G_N}+\bar{S}_{bulk,r}(\Sigma_{A|R}) ,~~ \partial \Sigma_{A|R}=A \cup \gamma_R^{>}(A)
\end{equation}
the constrained area at constant boost plus the bulk modular minimal entropy. $\bar{S}_{bulk,r}(\Sigma_{A|R})$ is the bulk entropy of the ``constrained homology hypersurface" $\Sigma_{A|R}$ (the region between the constrained surface and $A$) after evolving with the modular flow in the entanglement wedge of $R$, $r$.  Of course, whether one computes these quantum corrections from the bulk entropy of the dual of the modular evolved the state or by computing bulk modular minimal entropy in the original state, one should get the same answer.

In this way, classical and quantum contributions combine even if the bulk surface is not extremal, but the bulk quantum contribution is not simply a bulk entanglement entropy. So our work doesn't shed light on the definition of the ``generalized entropy" $A+S_{bulk}$ for non-extremal surfaces. Since that structure seems crucial for subregion-subregion duality \cite{Almheiri:2014lwa}, this doesn't give any evidence for subregion-subregion duality in the ``constrained entanglement wedge'' corresponding to $\gamma_{R}(A)$ (defined as the domain of dependence of $\Sigma_{A|R}$). 

This gives us a new perspective in the interpretation of $s_{min}$: in the bulk, the divergences of the entanglement entropy renormalize $G_N$. However, it seems unlikely that any renormalization scheme can get rid of divergences arising from kinks. In order for a bulk quantity to have a clear boundary interpretation, it should be bulk divergence free. In this way, the bulk modular minimal entropy should not have these extra divergences and thus $s_{min}$ can be interpreted as the local boost necessary to not have a kink along $\gamma_R(A)$. If the constrained surface $\gamma_R(A)$ doesn't have a constant boost angle along $\gamma(R)$, the naive bulk definition of the modular minimal entropy can't be made kink divergence free and thus there can not be a well defined boundary quantity dual to it.

As an example, consider the thermofield double state (fig \ref{fig:localTFD}c). The quantum corrections to the modular minimal entropy will be given by the entanglement entropy in the entanglement wedge of $A_t(-2 t_R)$ (which is bounded by the red surface). This is not the same as the entanglement entropy in the region between the constrained surface $\gamma_R(A_t)$ (purple surface) and $A_t$, this is easy to see because since $\gamma_R(A_t)$ has a kink, the bulk entanglement entropy between this surface and $A_t$ will have a divergence arising from the kink which is clearly not physical. In this case, the bulk modular minimal entropy is minimized by getting rid of the kink (since this gives rise to a bulk divergence) and thus we get the bulk entropy in the entanglement wedge of $A_t(-2 t_R)$.

\vspace{1cm}

\paragraph{Acknowledgments:}
We thank T.~Faulkner, D.~Marolf, and P.~Rath for useful discussions.  We would also like to thank the KITP for hospitality during the initial development of this work and the National Science Foundation for supporting the KITP under Grant No.\ PHY-1748958.  XD is supported in part by the National Science Foundation under Grant No.\ PHY-1820908 and by funds from the University of California.  AL acknowledges support from the Simons Foundation through the It from Qubit collaboration and would also like to thank the Department of Physics and Astronomy at the University of Pennsylvania for hospitality during the development of this work. XLQ is supported by the National Science Foundation under Grant No.\ 1720504 and by the David and Lucile Packard foundation.

\bibliography{bibliography}
\bibliographystyle{JHEP}
\end{document}